# Semiconservative replication in the quasispecies model II: Generalization to arbitrary lesion repair probabilities

Emmanuel Tannenbaum*
*Department of Chemistry and Chemical Biology, Harvard University, Cambridge, MA 02138*

James L. Sherley†
*Biological Engineering Division, Massachusetts Institute of Technology, Cambridge, MA 02139*

Eugene I. Shakhnovich‡
*Department of Chemistry and Chemical Biology,
Harvard University, Cambridge, Massachusetts 02138*

This paper extends the semiconservative quasispecies equations to account for arbitrary post-replication lesion repair efficiency. Such an extension could be an important tool for understanding processes such as cancer development and stem cell growth. Starting from the quasispecies dynamics over the space of genomes, we derive an equivalent dynamics over the space of ordered sequence pairs. From this set of equations, we are able to derive the infinite sequence length form of the dynamics for a class of "master-genome"-based fitness landscapes. We use these equations to solve for a "generalized" single-fitness-peak landscape, where the master genome can sustain a maximum number of lesions and remain viable. The central pattern that emerges from our studies is that imperfect lesion repair often leads to increased mutational robustness over semiconservative replication with completely efficient lesion repair. The reason for this is that imperfect lesion repair breaks some of the correlation between the parent and daughter strands, thereby preventing replication errors from destroying the information in the original genome. The result is a delayed error catastrophe over that expected from destroying the information in the original genome. In particular, we show that when only of the strands is necessary for conferring viability, then, when lesion repair is turned off, a semiconservatively replicating system becomes an effectively conservatively replicating one.

## I. INTRODUCTION

The quasispecies model has been a subject of ongoing research in the field of evolutionary dynamics for over three decades [1-22]. The model was originally introduced by Eigen in 1971 [1], as a way of accounting for the observed distribution of genotypes in evolution experiments with the Qβ RNA-virus [23]. It has since been applied to systems other than RNA genome evolution [18, 19, 24-30], and has even proven to give quantitative results in certain cases [18, 19].

The central result of the theory is the existence of an upper mutational threshold beyond which natural selection can no longer occur [1-3]. Below this threshold, a replicating population of genomes will eventually produce, over many generations, a "cloud" of closely related genomes clustered about one or a few fast replicating genomes. These "clouds" are termed quasispecies, and are characteristic of the evolutionary dynamics of many viruses, such as HIV [19, 31-33].

Above the mutational threshold, natural selection can no longer act to localize the population about the fast replicating genomes, and delocalization occurs over the entire genome space. This localization to delocalization transition is known as the error catastrophe [1-3], and it corresponds to the disappearance of any viable strains in the population. The error catastrophe has been observed experimentally [33, 34], and is believed to form the basis for a number of antiviral therapies [31-33].

In a recent paper [27], Tannenbaum *et al.* developed the quasispecies equations appropriate for describing DNA-based genomes. Such a description is a necessary first step toward making the quasispecies model a quantitative tool for analyzing the evolutionary dynamics of DNA-based life. The original quasispecies equations were developed to deal with the replication dynamics of single-stranded genomes, and hence assumed *conservative* replication. In conservative replication, the original genome is preserved by replication. Double-stranded DNA, by contrast, replicates *semiconservatively* [35]. In semiconservative replication, the original genome is not preserved by replication. Rather, the two strands of the genome separate, and each forms a template for the synthesis of the corresponding daughter strands [35]. Because errors can occur during the synthesis of the daughter strands, in principle the original genome is destroyed by the replication process, and it is possible that both daughter genomes will differ from the parent.

---
*Electronic address: etannenb@fas.harvard.edu
†Electronic address: jsherley@mit.edu
‡Electronic address: eugene@belok.harvard.edu

The semiconservative quasispecies equations in [27] were derived under the assumption of perfect lesion repair. Briefly, after replication has occurred, and both daughter genomes have been synthesized, it is possible that there are still mismatched base-pairs in the daughter genomes which were not corrected by various error-correcting mechanisms of the replication process itself (two such error-correcting mechanisms are the built-in proofreading capabilities of the DNA replicases, and the mismatch repair pathway) [35]. Any remaining mismatches will result in lesions along the DNA chain, which are recognized and repaired by various maintenance and repair enzymes present in the cell. However, after replication has occurred, it is no longer possible to distinguish between parent and daughter strands, and so the lesion is correctly repaired with a probability of 1/2.

In a recent work [30], Brumer and Shakhnovich studied the semiconservative quasispecies equations with imperfect lesion repair. The authors postulated that imperfect lesion repair may be necessary to reconcile the high point-mutation rates observed in certain cancers (the Microsatellite INstability, or MIN, tumors) with semiconservative replication. The argument stems from the fact that semiconservative replication is considerably less robust to the effect of replication errors than is conservative replication [27]. However, mutational robustness can be increased by reducing the efficiency of lesion repair. Imperfect lesion repair breaks the perfect correlation between the parent and daughter strands, thereby allowing for better preservation of genetic information. Thus, semiconservative replication with imperfect lesion repair can behave more like a conservatively replicating system in certain cases [30] (we will make this statement more precise later in the paper).

Subsequently, it was shown that imperfect lesion repair may also be necessary when modeling stem cell growth, in order to properly account for the effect of age-dependent chromosome segregation (known as the *immortal strand hypothesis*) [36]. Thus, it is apparent that imperfect lesion repair may be necessary for a proper modeling of the evolutionary dynamics of many biologically important phenomena.

Therefore, in this paper, we continue the work initiated in [27], and develop an extension of the semiconservative quasispecies equations which allows for arbitrary lesion repair probabilities. While the main results of this paper may be found in [37], the full details of the arbitrary lesion repair model are contained here.

This paper is organized as follows: In the following section, we present the finite genome length quasispecies equations for arbitrary lesion repair. While we cannot convert the dynamics over the space of double-stranded genomes to the space of single strands, as was possible in [27], we can nevertheless make an analogous transformation and convert the dynamics to the space of ordered strand pairs. In Section III, we go on to establish the infinite sequence length form of the equations for a class of fitness landscapes which are defined by a single, "master" genome. In Section IV, we explicitly solve for the equilibrium behavior of a subclass of these landscapes, which we call a generalized single-fitness-peak landscape. We also determine the critical mutation rate necessary for inducing error catastrophe for this class of fitness landscapes. In Section V, we explore the equilibrium behavior with specific examples, and discuss similarities and differences with both conservative and semiconservative replication with perfect lesion repair. We also present results from stochastic simulations of finite populations of replicating organisms, in order to corroborate the theory developed in this paper. Finally, in Section VI we conclude with a summary of our results and discuss plans for future research.

## II. THE FINITE SEQUENCE LENGTH EQUATIONS

### A. From double-stranded genomes to ordered sequence-pairs

Double-stranded DNA consists of two complementary, antiparallel strands [27, 35]. Each DNA genome is defined by the pair of strands $\{\sigma, \bar{\sigma}\} = \{\bar{\sigma}, \sigma\}$, where $\bar{\sigma}$ denotes the complement of $\sigma$. If each base is drawn from an alphabet of size $S$ (where $S = 4$ due to Watson-Crick pairing), and if $\bar{b}_i$ denotes the complement of a base $b_i$, then if $\sigma = b_1 \ldots b_L$, we have, by the antiparallel nature of DNA, that $\bar{\sigma} = \bar{b}_L \ldots \bar{b}_1$.

The replication of a DNA genome $\{\sigma, \bar{\sigma}\}$ may be divided into three stages:

1. Strand separation – The genome unzips to produce two parent strands, $\sigma$ and $\bar{\sigma}$.

2. Daughter strand synthesis – Each parent strand serves as the template for the synthesis of a complementary daughter strand.

3. Lesion repair after cell division.

An illustration of semiconservative replication may be found in [27].

This replication mechanism leads to the semiconservative quasispecies equations developed in [27]:

$$\frac{dx_{\{\sigma,\bar{\sigma}\}}}{dt} = -(\kappa_{\{\sigma,\bar{\sigma}\}} + \bar{\kappa}(t))x_{\{\sigma,\bar{\sigma}\}} + \sum_{\{\sigma',\bar{\sigma}'\}} \kappa_{\{\sigma',\bar{\sigma}'\}} x_{\{\sigma',\bar{\sigma}'\}} \times [p(\sigma', \{\sigma,\bar{\sigma}\}) + p(\bar{\sigma}', \{\sigma,\bar{\sigma}\})] \quad (1)$$

where $x_{\{\sigma,\bar{\sigma}\}}$ denotes the fraction of the population with genome $\{\sigma, \bar{\sigma}\}$, $\kappa_{\{\sigma,\bar{\sigma}\}}$ denotes the first-order growth rate constant, or fitness, associated with genome $\{\sigma, \bar{\sigma}\}$, $p(\sigma', \{\sigma, \bar{\sigma}\})$ denotes the probability that the parent strand $\sigma'$ forms the genome $\{\sigma, \bar{\sigma}\}$ after daughter strand synthesis and lesion repair, and $\bar{\kappa}(t) \equiv \sum_{\{\sigma,\bar{\sigma}\}} \kappa_{\{\sigma,\bar{\sigma}\}} x_{\{\sigma,\bar{\sigma}\}}$ is the mean fitness of the population.

When lesion repair is imperfect, the correlation between the two strands is broken, and we must consider a more generalized dynamics over genomes of the form $\{\sigma, \sigma'\}$, where both $\sigma$ and $\sigma'$ are arbitrary. Following the derivation in [27], we obtain the quasispecies equations,

$$\frac{dx_{\{\sigma,\sigma'\}}}{dt} = -(\kappa_{\{\sigma,\sigma'\}} + \bar{\kappa}(t))x_{\{\sigma,\sigma'\}}$$
$$+ \sum_{\{\sigma'',\sigma'''\}} \kappa_{\{\sigma'',\sigma'''\}} x_{\{\sigma'',\sigma'''\}} \times$$
$$[p((\sigma'', \sigma'''), \{\sigma, \sigma'\}) + p((\sigma''', \sigma''), \{\sigma, \sigma'\})] \quad (2)$$

Here, $p((\sigma'', \sigma'''), \{\sigma, \sigma'\})$ denotes the probability that parent strand $\sigma''$, as part of genome $\{\sigma'', \sigma'''\}$, becomes genome $\{\sigma, \sigma'\}$ after daughter strand synthesis and lesion repair. In addition, we have $\bar{\kappa}(t) = \sum_{\{\sigma'',\sigma'''\}} \kappa_{\{\sigma'',\sigma'''\}} x_{\{\sigma'',\sigma'''\}}$. The definitions are otherwise unchanged from the original semiconservative equations.

In the semiconservative quasispecies equations, the complementarity property allows one to convert the quasispecies dynamics over the space of double-stranded genomes into an equivalent, and considerably simpler, dynamics over the space of single strands [27]. With imperfect lesion repair, the lack of perfect correlation between the two strands in the genome makes a conversion to a single strand model impossible. Nevertheless, we can make an analogous transformation of the dynamics, from double-stranded genomes $\{\sigma, \sigma'\}$ to *ordered pairs* of strands, $(\sigma, \sigma')$, as follows: We define $y_{(\sigma,\sigma')} = y_{(\sigma',\sigma)} = \frac{1}{2} x_{\{\sigma,\sigma'\}}$ if $\sigma \neq \sigma'$, and $y_{(\sigma,\sigma)} = x_{\{\sigma,\sigma\}}$. Also, we define $\kappa_{(\sigma,\sigma')} = \kappa_{(\sigma',\sigma)} = \kappa_{\{\sigma,\sigma'\}}$. We then have that,

$$\sum_{\{\sigma'',\sigma'''\}} \kappa_{\{\sigma'',\sigma'''\}} x_{\{\sigma'',\sigma'''\}} \times$$
$$[p((\sigma'', \sigma'''), \{\sigma, \sigma'\}) + p((\sigma''', \sigma''), \{\sigma, \sigma'\})]$$
$$= 2 \sum_{\{\sigma'',\sigma'''\}, \sigma'' \neq \sigma'''} [\kappa_{(\sigma'',\sigma''')} y_{(\sigma'',\sigma''')} p((\sigma'', \sigma'''), \{\sigma, \sigma'\})$$
$$+ \kappa_{(\sigma''',\sigma'')} y_{(\sigma''',\sigma'')} p((\sigma''', \sigma''), \{\sigma, \sigma'\})]$$
$$+ 2 \sum_{\{\sigma'',\sigma''\}} \kappa_{(\sigma'',\sigma'')} y_{(\sigma'',\sigma'')} p((\sigma'', \sigma''), \{\sigma, \sigma'\})$$
$$= 2 \sum_{(\sigma'',\sigma''')} \kappa_{(\sigma'',\sigma''')} y_{(\sigma'',\sigma''')} p((\sigma'', \sigma'''), \{\sigma, \sigma'\}) \quad (3)$$

Finally, we define $p((\sigma'', \sigma'''), (\sigma, \sigma'))$ to be the probability that $\sigma''$, as part of genome $\{\sigma'', \sigma'''\}$, becomes $\sigma$, with daughter strand $\sigma'$ (after daughter strand synthesis and lesion repair). Then it follows that,

$$p((\sigma'',\sigma'''), \{\sigma,\sigma'\}) = \begin{cases} p((\sigma'',\sigma'''),(\sigma,\sigma')) + p((\sigma'',\sigma'''),(\sigma',\sigma)) & \text{if } \sigma \neq \sigma' \\ p((\sigma'',\sigma'''),(\sigma,\sigma')) & \text{if } \sigma = \sigma' \end{cases} \quad (4)$$

For $\sigma' \neq \sigma$, we therefore obtain that,

$$\frac{dy_{(\sigma,\sigma')}}{dt} = \frac{1}{2} \frac{dx_{\{\sigma,\sigma'\}}}{dt}$$
$$= -(\kappa_{(\sigma,\sigma')} + \bar{\kappa}(t))y_{(\sigma,\sigma')}$$
$$+ \sum_{(\sigma'',\sigma''')} \kappa_{(\sigma'',\sigma''')} y_{(\sigma'',\sigma''')} \times$$
$$[p((\sigma'', \sigma'''), (\sigma, \sigma')) + p((\sigma'', \sigma'''), (\sigma', \sigma))] \quad (5)$$

The same equation holds for $y_{(\sigma,\sigma)}$, since $2p((\sigma'',\sigma'''), \{\sigma,\sigma\}) = p((\sigma'',\sigma'''),(\sigma,\sigma)) + p((\sigma'',\sigma'''),(\sigma,\sigma))$. Therefore, the quasispecies dynamics over the space of ordered sequence-pairs is given by,

$$\frac{dy_{(\sigma,\sigma')}}{dt} = -(\kappa_{(\sigma,\sigma')} + \bar{\kappa}(t))y_{(\sigma,\sigma')}$$
$$+ \sum_{(\sigma'',\sigma''')} \kappa_{(\sigma'',\sigma''')} y_{(\sigma'',\sigma''')} \times$$
$$[p((\sigma'', \sigma'''), (\sigma, \sigma')) + p((\sigma'', \sigma'''), (\sigma', \sigma))] \quad (6)$$

As a final derivation in this subsection, we will obtain an equivalent formulation of Eq. (6) which will prove useful later. To begin, suppose that the fitness landscape is such that $\kappa_{(\bar{\sigma},\bar{\sigma}')} = \kappa_{(\sigma,\sigma')}$. Furthermore, suppose we have that $p((\bar{\sigma}'', \bar{\sigma}'''), (\bar{\sigma}, \bar{\sigma}')) = p((\sigma'', \sigma'''), (\sigma, \sigma'))$. Then, if our population is initially lesion-free, we claim that $y_{(\bar{\sigma},\bar{\sigma}')} = y_{(\sigma,\sigma')}$ at all times.

To see this, note first that a lesion-free population is equivalent to the statement that $y_{(\sigma,\sigma')} = 0$ if $\sigma' \neq \bar{\sigma}$. Then if $\sigma' \neq \bar{\sigma}$, it certainly follows that $\bar{\sigma}' \neq \bar{\bar{\sigma}}$, hence $y_{(\bar{\sigma},\bar{\sigma}')} = 0 = y_{(\sigma,\sigma')}$. On the other hand, if $\sigma' = \bar{\sigma}$, then $y_{(\bar{\sigma},\bar{\sigma}')} = y_{(\sigma',\sigma)} = y_{(\sigma,\sigma')}$. Therefore, a population which is lesion-free satisfies the property that $y_{(\bar{\sigma},\bar{\sigma}')} = y_{(\sigma,\sigma')}$ for all sequence pairs $(\sigma, \sigma')$.

Then in order to prove that $y_{(\bar{\sigma},\bar{\sigma}')} = y_{(\sigma,\sigma')}$ at all times, we need only show that $y_{(\bar{\sigma},\bar{\sigma}')} = y_{(\sigma,\sigma')}$ at some time $t$ implies that $dy_{(\bar{\sigma},\bar{\sigma}')}/dt = dy_{(\sigma,\sigma')}/dt$. We have,

$$\begin{aligned}
\frac{dy_{(\bar{\sigma},\bar{\sigma}')}}{dt} &= -(\kappa_{(\bar{\sigma},\bar{\sigma}')} + \bar{\kappa}(t))y_{(\bar{\sigma},\bar{\sigma}')} \\
&+ \sum_{(\sigma'',\sigma''')} \kappa_{(\sigma'',\sigma''')} y_{(\sigma'',\sigma''')} \times \\
&\quad [p((\sigma'',\sigma'''),(\bar{\sigma},\bar{\sigma}')) + p((\sigma'',\sigma'''),(\bar{\sigma}',\bar{\sigma}))] \\
&= -(\kappa_{(\sigma,\sigma')} + \bar{\kappa}(t))y_{(\sigma,\sigma')} \\
&+ \sum_{(\sigma'',\sigma''')} \kappa_{(\bar{\sigma}'',\bar{\sigma}''')} y_{(\bar{\sigma}'',\bar{\sigma}''')} \times \\
&\quad [p((\bar{\sigma}'',\bar{\sigma}'''),(\bar{\sigma},\bar{\sigma}')) + p((\bar{\sigma}'',\bar{\sigma}'''),(\bar{\sigma}',\bar{\sigma}))] \\
&= -(\kappa_{(\sigma,\sigma')} + \bar{\kappa}(t))y_{(\sigma,\sigma')} \\
&+ \sum_{(\sigma'',\sigma''')} \kappa_{(\sigma'',\sigma''')} y_{(\sigma'',\sigma''')} \times \\
&\quad [p((\sigma'',\sigma'''),(\sigma,\sigma')) + p((\sigma'',\sigma'''),(\sigma',\sigma))] \\
&= \frac{dy_{(\sigma,\sigma')}}{dt} \quad (7)
\end{aligned}$$

which establishes our claim.

So let us assume that our fitness landscape is such that $\kappa_{(\bar{\sigma},\bar{\sigma}')} = \kappa_{(\sigma,\sigma')}$, and also that $p((\bar{\sigma}'',\bar{\sigma}'''),(\bar{\sigma},\bar{\sigma}')) = p((\sigma'',\sigma'''),(\sigma,\sigma'))$. If our population is initially lesion-free, then for all sequence pairs $(\sigma,\sigma')$ we have $y_{(\bar{\sigma},\bar{\sigma}')} = y_{(\sigma,\sigma')}$. This gives,

$$\begin{aligned}
\frac{dy_{(\sigma,\sigma')}}{dt} &= -(\kappa_{(\sigma,\sigma')} + \bar{\kappa}(t))y_{(\sigma,\sigma')} \\
&+ \sum_{(\sigma'',\sigma''')} \kappa_{(\sigma'',\sigma''')} y_{(\sigma'',\sigma''')} p((\sigma'',\sigma'''),(\sigma,\sigma')) \\
&+ \sum_{(\sigma'',\sigma''')} \kappa_{(\bar{\sigma}'',\bar{\sigma}''')} y_{(\bar{\sigma}'',\bar{\sigma}''')} p((\bar{\sigma}'',\bar{\sigma}'''),(\sigma',\sigma)) \\
&= -(\kappa_{(\sigma,\sigma')} + \bar{\kappa}(t))y_{(\sigma,\sigma')} \\
&+ \sum_{(\sigma'',\sigma''')} \kappa_{(\sigma'',\sigma''')} y_{(\sigma'',\sigma''')} p((\sigma'',\sigma'''),(\sigma,\sigma')) \\
&+ \sum_{(\sigma'',\sigma''')} \kappa_{(\sigma'',\sigma''')} y_{(\sigma'',\sigma''')} p((\sigma'',\sigma'''),(\bar{\sigma}',\bar{\sigma})) \quad (8)
\end{aligned}$$

which can be simplified to give,

$$\begin{aligned}
\frac{dy_{(\sigma,\sigma')}}{dt} &= -(\kappa_{(\sigma,\sigma')} + \bar{\kappa}(t))y_{(\sigma,\sigma')} \\
&+ \sum_{(\sigma'',\sigma''')} \kappa_{(\sigma'',\sigma''')} y_{(\sigma'',\sigma''')} \times \\
&\quad [p((\sigma'',\sigma'''),(\sigma,\sigma')) + p((\sigma'',\sigma'''),(\bar{\sigma}',\bar{\sigma}))] \quad (9)
\end{aligned}$$

We will make use of these equations when considering the behavior of the quasispecies dynamics in the limit of infinite genome lengths.

### B. Determination of $p((\sigma'',\sigma'''),(\sigma,\sigma'))$

We now compute $p((\sigma'',\sigma'''),(\sigma,\sigma'))$, assuming that with each genome $\{\sigma,\sigma'\}$ there is a base-pair independent mismatch probability, denoted by $\epsilon_{\{\sigma,\sigma'\}}$, and a base-pair independent lesion repair probability, denoted by $\lambda_{\{\sigma,\sigma'\}}$ (the genome dependence of the mismatch and lesion repair probabilities arises from the fact that different genomes may code for different enzymes, or none at all, that are involved in DNA repair. See for instance [24–26]).

We begin with some definitions: Define $\sigma_C$ to be the subsequence of bases in $\sigma$ which are complementary with the corresponding bases in $\sigma'$. That is, suppose $\sigma = b_1 \ldots b_L$, and suppose for indices $i_1 < i_2 < \cdots < i_k$ we have that $\bar{b}_{i_j} = b'_{L-i_j+1}$. Then $\sigma_C = b_{i_1} \ldots b_{i_k}$. We also define $\sigma'_C$ to be the subsequence of corresponding bases in $\sigma'$, so that $\sigma'_C = b'_{L-i_k+1} \ldots b'_{L-i_1+1}$. Finally, let $\sigma''_C$ denote the subsequence of bases in $\sigma''$ corresponding to the bases in $\sigma_C$, so that $\sigma''_C = b''_{i_1} \ldots b''_{i_k}$.

Now, define $\sigma_{NC}$ to be the subsequence of bases in $\sigma$ which are not complementary with the corresponding bases in $\sigma'$. That is, given the complementary indices $i_1 < i_2 < \cdots < i_k$ defined above, let $i'_1 < i'_2 < \cdots < i'_{L-k}$ be the remaining indices. Then $\sigma_{NC} = b_{i'_1} \ldots b_{i'_{L-k}}$. We define $\sigma'_{NC}$ to be the subsequence of corresponding bases in $\sigma'$, so that $\sigma'_{NC} = b'_{L-i'_{L-k}+1} \ldots b'_{L-i'_1+1}$. Finally, we let $\sigma''_{NC}$ denote the subsequence of bases in $\sigma''$ corresponding to the bases in $\sigma_{NC}$, so that $\sigma''_{NC} = b''_{i'_1} \ldots b''_{i'_{L-k}}$.

We now let $p((\sigma'',\sigma'''');\sigma''')$ denote the probability that $\sigma''$, as part of genome $\{\sigma'',\sigma'''\}$, is paired with $\sigma''''$ during daughter strand synthesis. We also let $p((\sigma'',\sigma'''') \to (\sigma,\sigma');\sigma''')$ denote the probability that $\sigma''$ becomes $\sigma$ and $\sigma''''$ becomes $\sigma'$ during lesion repair (the presence of the $\sigma'''$ in this notation is to indicate that $\sigma''$ comes from genome $\{\sigma'',\sigma'''\}$. Presumably, the enzymes involved in lesion repair are the ones that came from the original parent cell, hence the lesion repair probability should be $\lambda_{\{\sigma'',\sigma'''\}}$). Then we have that,

$$\begin{aligned}
p((\sigma'',\sigma'''),(\sigma,\sigma')) &= \sum_{\sigma''''} p((\sigma'',\sigma'''');\sigma''') \times \\
&\quad p((\sigma'',\sigma'''') \to (\sigma,\sigma');\sigma''') \quad (10)
\end{aligned}$$

Consider some base $b''_i$ in $\sigma''$, and suppose that $b''_i$ is part of $\sigma''_C$. If $b''_i$ differs from the corresponding base $b_i$ in $\sigma_C$, then it is clear that during daughter strand synthesis, it must be paired with $\bar{b}_i$, and during lesion repair it is $b''_i$ that must be repaired to form $b_i$. Therefore, if $l_C \equiv D_H(\sigma''_C, \sigma_C)$ denotes the Hamming distance between $\sigma''_C$ and $\sigma_C$, then $b''_i$ must be paired with $\bar{b}_i$, and the $(b''_i, \bar{b}_i)$ lesion must be repaired to $(b_i, \bar{b}_i)$, in $l_C$ places. The probability of mispairing a given $b''_i$ with $\bar{b}_i$ is $\epsilon_{\{\sigma'',\sigma'''\}}/(S-1)$. The probability of lesion repair is $\lambda_{\{\sigma'',\sigma'''\}}$. Finally, assuming lesion repair occurs, the probability of repairing $b''_i$ is $1/2$. Assuming that $\sigma''''$ is chosen to satisfy the pairing requirements described above, we obtain a factor of $(\lambda_{\{\sigma'',\sigma'''\}}\epsilon_{\{\sigma'',\sigma'''\}}/2(S-1))^{l_C}$ contribution to $p((\sigma'',\sigma'''');\sigma''')p((\sigma'',\sigma'''') \to (\sigma,\sigma');\sigma''')$.

Now, let $L_C$ denote the length of $\sigma_C$, so that $\sigma_C''$ and $\sigma_C$ are equal in $L_C - l_C$ positions. Then, given some $b_i''$ in one of these $L_C - l_C$ positions, it can be paired with any other base. Let $l_{C,1}$ denote the number, among these positions, where $b_i''$ is mispaired with a base other than $\bar{b}_i'' = \bar{b}_i$. Then, among these $L_C - l_C$ positions, $b_i''$ is paired with $\bar{b}_i''$ in $L_C - l_C - l_{C,1}$ positions. Since lesion repair must happen in $l_{C,1}$ positions, then for an appropriately chosen $\sigma''''$, we have a factor of $(\lambda_{\{\sigma'',\sigma'''\}}\epsilon_{\{\sigma'',\sigma'''\}}/2(S-1))^{l_{C,1}}(1-\epsilon_{\{\sigma'',\sigma'''\}})^{L_C - l_C - l_{C,1}}$ contribution to $p((\sigma'',\sigma'''');\sigma''')p((\sigma'',\sigma'''') \to (\sigma,\sigma');\sigma''')$.

Finally, let $L_{NC}$ denote the length of $\sigma_{NC}$. Since $\sigma_{NC}$ and $\sigma_{NC}'$ are not complementary, no lesion repair can happen at positions in $\sigma_{NC}''$. Therefore, $\sigma_{NC}''$ cannot be changed, hence we must have $\sigma_{NC}'' = \sigma_{NC}$. Also, a mismatch must occur at all sites along $\sigma_{NC}''$ to form the corresponding bases in $\sigma_{NC}'$. Once again, for an appropriately chosen $\sigma''''$, we have a factor of $\delta_{\sigma_{NC}''\sigma_{NC}}((1-\lambda_{\{\sigma'',\sigma'''\}})\epsilon_{\{\sigma'',\sigma'''\}}/(S-1))^{L_{NC}}$ contribution to $p((\sigma'',\sigma'''');\sigma''')p((\sigma'',\sigma'''') \to (\sigma,\sigma');\sigma''')$. Therefore, given a daughter strand $\sigma''''$ for which $(\sigma'',\sigma'''')$ can become $(\sigma,\sigma')$ after lesion repair, we have,

$$p((\sigma'',\sigma'''');\sigma''')p((\sigma'',\sigma'''') \to (\sigma,\sigma');\sigma''') =$$
$$\delta_{\sigma_{NC}''\sigma_{NC}}(\frac{\lambda\epsilon_{\{\sigma'',\sigma'''\}}}{2(S-1)})^{l_C}(\frac{(1-\lambda)\epsilon_{\{\sigma'',\sigma'''\}}}{S-1})^{L_{NC}} \times$$
$$(\frac{\lambda\epsilon_{\{\sigma'',\sigma'''\}}}{2(S-1)})^{l_{C,1}}(1-\epsilon_{\{\sigma'',\sigma'''\}})^{L_C - l_C - l_{C,1}} \quad (11)$$

To evaluate the sum in Eq. (11), we need only sum over those $\sigma''''$ for which $p((\sigma'',\sigma'''');\sigma''')p((\sigma'',\sigma'''') \to (\sigma,\sigma');\sigma''')$ is nonzero. Thus, we sum over all possible values of $l_{C,1}$, taking into account degeneracies for each value of $l_{C,1}$. This gives,

$$\begin{aligned}
p((\sigma'',\sigma'''),(\sigma,\sigma')) &= \sum_{\sigma''''} p((\sigma'',\sigma'''');\sigma''')p((\sigma'',\sigma'''') \to (\sigma,\sigma');\sigma''') \\
&= \delta_{\sigma_{NC}''\sigma_{NC}}(\frac{\lambda_{\{\sigma'',\sigma'''\}}\epsilon_{\{\sigma'',\sigma'''\}}}{2(S-1)})^{l_C}(\frac{(1-\lambda_{\{\sigma'',\sigma'''\}})\epsilon_{\{\sigma'',\sigma'''\}}}{S-1})^{L_{NC}} \times \\
&\quad \sum_{l_{C,1}=0}^{L_C-l_C} \binom{L_C - l_C}{l_{C,1}}(S-1)^{l_{C,1}}(\frac{\lambda_{\{\sigma'',\sigma'''\}}\epsilon_{\{\sigma'',\sigma'''\}}}{2(S-1)})^{l_{C,1}}(1-\epsilon_{\{\sigma'',\sigma'''\}})^{L_C-l_C-l_{C,1}} \\
&= \delta_{\sigma_{NC}''\sigma_{NC}}(\frac{\lambda_{\{\sigma'',\sigma'''\}}\epsilon_{\{\sigma'',\sigma'''\}}}{2(S-1)})^{l_C}(\frac{(1-\lambda_{\{\sigma'',\sigma'''\}})\epsilon_{\{\sigma'',\sigma'''\}}}{S-1})^{L_{NC}}(1-\epsilon_{\{\sigma'',\sigma'''\}}(1-\frac{\lambda_{\{\sigma'',\sigma'''\}}}{2}))^{L_C-l_C}
\end{aligned} \quad (12)$$

Note that $L_C = L - L_{NC}$, and note that since $L_{NC}$ is simply the number of positions where $\sigma$ and $\sigma'$ are not complementary, it follows that $L_{NC} = D_H(\sigma,\bar{\sigma}')$. Therefore, our final formula is,

$$\begin{aligned}
p((\sigma'',\sigma'''),(\sigma,\sigma')) &= \delta_{\sigma_{NC}''\sigma_{NC}}(\frac{\lambda_{\{\sigma'',\sigma'''\}}\epsilon_{\{\sigma'',\sigma'''\}}}{2(S-1)})^{D_H(\sigma_C'',\sigma_C)}(\frac{(1-\lambda_{\{\sigma'',\sigma'''\}})\epsilon_{\{\sigma'',\sigma'''\}}}{S-1})^{D_H(\sigma,\bar{\sigma}')} \times \\
&\quad (1-\epsilon_{\{\sigma'',\sigma'''\}}(1-\frac{\lambda_{\{\sigma'',\sigma'''\}}}{2}))^{L-D_H(\sigma,\bar{\sigma}')-D_H(\sigma_C'',\sigma_C)}
\end{aligned} \quad (13)$$

For the remainder of this paper, we will assume that $\epsilon_{\{\sigma,\sigma'\}}$ and $\lambda_{\{\sigma,\sigma'\}}$ are genome independent, and hence may be denoted by $\epsilon$ and $\lambda$ (unless otherwise indicated).

### C. Obtaining the $\lambda = 1$ semiconservative equations

When $\lambda = 1$, it follows that $p((\sigma'',\sigma'''),(\sigma,\sigma')) = \delta_{\bar{\sigma}\sigma'}p((\sigma'',\sigma'''),(\sigma,\bar{\sigma}))$, since, with perfect lesion repair, all post-replication lesions are removed. Therefore, if $\sigma' \neq \bar{\sigma}$, then,

$$\frac{dy_{(\sigma,\sigma')}}{dt} = -(\kappa_{(\sigma,\sigma')} + \bar{\kappa}(t))y_{(\sigma,\sigma')} \quad (14)$$

This implies that genomes with lesions will eventually disappear from the population. Furthermore, if an initial population of genomes is lesion-free, then no lesions will appear in the population, hence in such a case we may take $y_{(\sigma,\sigma')} = 0$ for $\sigma' \neq \bar{\sigma}$, and restrict our dynamics to the space of complementary ordered sequence-pairs, denoted $\{(\sigma, \bar{\sigma})\}$. We then have,

$$\begin{aligned}\frac{dy_{(\sigma,\bar{\sigma})}}{dt} &= -(\kappa_{(\sigma,\bar{\sigma})} + \bar{\kappa}(t))y_{(\sigma,\bar{\sigma})} \\ &+ \sum_{\sigma'} \kappa_{(\sigma',\bar{\sigma}')} y_{(\sigma',\bar{\sigma}')} \times \\ &\quad [p((\sigma',\bar{\sigma}'),(\sigma,\bar{\sigma})) + p((\sigma',\bar{\sigma}'),(\bar{\sigma},\sigma))]\end{aligned}$$
(15)

Now, for $\lambda = 1$ note that,

$$\begin{aligned}p((\sigma',\bar{\sigma}'),(\sigma,\bar{\sigma})) &= (\frac{\epsilon_{(\sigma',\bar{\sigma}')}}{2(S-1)})^{D_H(\sigma',\sigma)} \times \\ &\quad (1 - \frac{\epsilon_{(\sigma',\bar{\sigma}')}}{2})^{L-D_H(\sigma',\sigma)}\end{aligned}$$
(16)

Note then that since $D_H(\bar{\sigma}', \bar{\sigma}) = D_H(\sigma', \sigma)$, we have that $p((\bar{\sigma}', \sigma'),(\bar{\sigma}, \sigma)) = p((\sigma', \bar{\sigma}'),(\sigma, \bar{\sigma}))$, and so,

$$\begin{aligned}\frac{dy_{(\sigma,\bar{\sigma})}}{dt} &= -(\kappa_{(\sigma,\bar{\sigma})} + \bar{\kappa}(t))y_{(\sigma,\bar{\sigma})} \\ &+ \sum_{\sigma'} \kappa_{(\sigma',\bar{\sigma}')} y_{(\sigma',\bar{\sigma}')} p((\sigma',\bar{\sigma}'),(\sigma,\bar{\sigma})) \\ &+ \sum_{\sigma'} \kappa_{(\bar{\sigma}',\sigma')} y_{(\bar{\sigma}',\sigma')} p((\bar{\sigma}',\sigma'),(\bar{\sigma},\sigma)) \\ &= -(\kappa_{(\sigma,\bar{\sigma})} + \bar{\kappa}(t))y_{(\sigma,\bar{\sigma})} \\ &+ 2 \sum_{\sigma'} \kappa_{(\sigma',\bar{\sigma}')} y_{(\sigma',\bar{\sigma}')} p((\sigma',\bar{\sigma}'),(\sigma,\bar{\sigma}))\end{aligned}$$
(17)

Defining $y_\sigma = y_{\{\sigma,\bar{\sigma}\}}$, $\epsilon_\sigma = \epsilon_{(\sigma,\bar{\sigma})}$, and $\kappa_\sigma = \kappa_{(\sigma,\bar{\sigma})}$ gives,

$$\begin{aligned}\frac{dy_\sigma}{dt} &= -(\kappa_\sigma + \bar{\kappa}(t))y_\sigma \\ &+ 2 \sum_{\sigma'} \kappa_{\sigma'} y_{\sigma'} (\frac{\epsilon_{\sigma'}}{2(S-1)})^{D_H(\sigma',\sigma)} (1 - \frac{\epsilon_{\sigma'}}{2})^{L-D_H(\sigma',\sigma)}\end{aligned}$$
(18)

which are exactly the original semiconservative equations derived in [27].

### D. $\lambda = 0$ equations

Before concluding this section, we will show that when lesion repair is turned off, then the semiconservative quasispecies equations can be transformed into equations which are similar in form to the conservative quasispecies equations. This is essentially a rederivation of a result of Brumer and Shakhnovich [30], done with our sequence-pair formalism.

For this derivation, we make the assumption that $\epsilon_{\{\sigma,\sigma'\}}$ is a constant $\epsilon$ for all genomes. This implies that $p((\sigma'', \sigma'''); \sigma''')$ does not depend on $\sigma'''$, hence the term may be dropped from the notation. Since lesion repair is turned off, we have $p((\sigma'', \sigma'''),(\sigma, \sigma')) = \delta_{\sigma''\sigma} p(\sigma, \sigma')$. This gives,

$$\begin{aligned}\frac{dy_{(\sigma,\sigma')}}{dt} &= -(\kappa_{(\sigma,\sigma')} + \bar{\kappa}(t))y_{(\sigma,\sigma')} \\ &+ \sum_{\sigma''} \kappa_{(\sigma,\sigma'')} y_{(\sigma,\sigma'')} p(\sigma,\sigma') \\ &+ \sum_{\sigma''} \kappa_{(\sigma',\sigma'')} y_{(\sigma',\sigma'')} p(\sigma',\sigma)\end{aligned}$$
(19)

Now, define $y_\sigma = \sum_{\sigma'} y_{(\sigma,\sigma')}$. Also, define $\kappa_\sigma = \sum_{\sigma'} \kappa_{(\sigma,\sigma')} y_{(\sigma,\sigma')} / y_\sigma$. We then have,

$$\begin{aligned}\frac{dy_\sigma}{dt} &= -(\kappa_\sigma + \bar{\kappa}(t))y_\sigma \\ &+ \kappa_\sigma y_\sigma \\ &+ \sum_{\sigma'} \kappa_{\sigma'} y_{\sigma'} p(\sigma',\sigma) \\ &= \sum_{\sigma'} \kappa_{\sigma'} y_{\sigma'} p(\sigma',\sigma) - \bar{\kappa}(t) y_\sigma\end{aligned}$$
(20)

where we have used the fact that $\sum_{\sigma'} p(\sigma, \sigma') = 1$.

Note that we have transformed the semiconservative quasispecies equations into a set of equations that look like the conservative equations, the key difference being that the fitnesses $\kappa_\sigma$ are concentration-dependent. However, it is possible to show that when the fitness depends on only one of the strands, then the conservative equations are obtained exactly (this will be done later in the paper).

### III. THE "MASTER" GENOME FITNESS LANDSCAPE

#### A. Infinite sequence length equations

We will now develop the infinite sequence length equations for a class of fitness landscapes defined by what we call a "master" genome $\{\sigma_0, \bar{\sigma}_0\}$. A subclass of these landscapes is a generalization of the single-fitness-peak landscape [3, 27], which is the simplest landscape for which analytical results are obtainable. We will solve for the equilibrium mean fitness and the error threshold associated with this class of landscapes in the next section.

Before proceeding, we note that the infinite sequence length equations are taken with $\mu \equiv L\epsilon$ held constant. Because the probability of correct daughter strand synthesis is $(1 - \epsilon)^L \to e^{-\mu}$ as $L \to \infty$, holding $\mu$ constant amounts to fixing the genome replication fidelity in the limit of infinite sequence length.

The "master" genome $\{\sigma_0, \bar{\sigma}_0\}$ gives rise to the ordered sequence pairs $(\sigma_0, \bar{\sigma}_0)$ and $(\bar{\sigma}_0, \sigma_0)$. In the limit of infinite sequence length, it is possible to show that, with probability 1, the sequences $\sigma_0$ and $\bar{\sigma}_0$ become infinitely separated from each other, i.e. $D_H(\sigma_0, \bar{\sigma}_0) \to \infty$ [27]. Thus, we may regard $(\sigma_0, \bar{\sigma}_0)$ and $(\bar{\sigma}_0, \sigma_0)$ as infinitely separated from each other in the ordered sequence pair space.

The infinite separation between $\sigma_0$ and $\bar{\sigma}_0$ allows a division of the sequence pairs into three classes. A sequence pair $(\sigma, \sigma')$ is said to be of the *first class* if $D_H(\sigma, \sigma_0)$ and $D_H(\sigma', \bar{\sigma}_0)$ are both finite. A sequence pair $(\sigma, \sigma')$ is said to be of the *second class* if $D_H(\sigma, \bar{\sigma}_0)$ and $D_H(\sigma', \sigma_0)$ are both finite. Finally, a sequence pair not belonging to either one of the first two classes is said to belong to the *third class*. Using the Triangle Inequality, it is readily shown that a sequence pair cannot belong to more than one class.

A given sequence pair $(\sigma, \sigma')$ of the first class can be characterized by four parameters, denoted $l_C, l_L, l_R,$ and $l_B$. The first parameter, $l_C$, denotes the number of positions where $\sigma$ and $\sigma'$ are complementary, yet differ from the corresponding positions in $\sigma_0$ and $\bar{\sigma}_0$, respectively. The second parameter, $l_L$, denotes the number of positions where $\sigma$ differs from $\sigma_0$, but the complementary positions in $\sigma'$ are equal to the corresponding ones in $\bar{\sigma}_0$. The third parameter, $l_R$, denotes the number of positions where $\sigma$ is equal to the ones in $\sigma_0$, but the complementary positions in $\sigma'$ differ from the corresponding ones in $\bar{\sigma}_0$. Finally, the fourth parameter, $l_B$, denotes the number of positions where $\sigma$ and $\sigma'$ are not complementary, and also differ from the corresponding positions in $\sigma_0$ and $\bar{\sigma}_0$, respectively. A sequence pair $(\sigma, \sigma')$ of the second class may be similarly characterized (except $\sigma_0$ and $\bar{\sigma}_0$ are swapped in the definitions given above).

We assume that the fitness of a given sequence pair of the first class is determined by $l_C$, $l_L$, $l_R$, and $l_B$, hence we may write that $\kappa_{(\sigma, \sigma')} = \kappa_{(l_C, l_L, l_R, l_B)}$. The fitness of a sequence pair $(\sigma, \sigma')$ of the second class is determined by noting that $(\sigma', \sigma)$ is of the first class, and that $\kappa_{(\sigma, \sigma')} = \kappa_{(\sigma', \sigma)}$. We take the third class sequence pairs to be unviable, with a first-order growth rate of 1.

We also assume that $\kappa_{(l_C, l_L, l_R, l_B)} = \kappa_{(l_C, l_R, l_L, l_B)}$. This is a natural assumption to make if one assumes symmetry between the two master strands $\sigma_0$ and $\bar{\sigma}_0$. This assumption also implies that $\kappa_{(\bar{\sigma}, \bar{\sigma}')} = \kappa_{(\sigma, \sigma')}$. To see this, let us first suppose that $(\sigma, \sigma')$ is of the first class, and is characterized by the parameters $l_C, l_L, l_R,$ and $l_B$. Then $(\bar{\sigma}', \bar{\sigma})$ is also of the first class. Because taking the complement of a sequence essentially amounts to a relabelling of the bases defined by a one-to-one map, and to a reversal in the sequence direction, it follows that $(\bar{\sigma}, \bar{\sigma}')$ is a sequence pair of the second class, characterized by the parameters $l_C, l_L, l_R,$ and $l_B$. Therefore, $(\bar{\sigma}', \bar{\sigma})$ is characterized by the parameters $l_C, l_R, l_L,$ and $l_B$, and so $\kappa_{(\bar{\sigma}, \bar{\sigma}')} = \kappa_{(\bar{\sigma}', \bar{\sigma})} = \kappa_{(l_C, l_R, l_L, l_B)} = \kappa_{(l_C, l_L, l_R, l_B)} = \kappa_{(\sigma, \sigma')}$.

If $(\sigma, \sigma')$ is of the second class, then $(\sigma', \sigma)$ is of the first class. We then have $\kappa_{(\bar{\sigma}, \bar{\sigma}')} = \kappa_{(\bar{\sigma}', \bar{\sigma})} = \kappa_{(\sigma', \sigma)} = \kappa_{(\sigma, \sigma')}$.

Finally, if $(\sigma, \sigma')$ is of the third class, then using the identity $D_H(\bar{\sigma}_1, \bar{\sigma}_2) = D_H(\sigma_1, \sigma_2)$ we can show that $(\bar{\sigma}, \bar{\sigma}')$ is also of the third class. Therefore, $\kappa_{(\bar{\sigma}, \bar{\sigma}')} = 1 = \kappa_{(\sigma, \sigma')}$.

Based on our formula for $p((\sigma'', \sigma'''), (\sigma, \sigma'))$, we have that $p((\bar{\sigma}'', \bar{\sigma}'''), (\bar{\sigma}, \bar{\sigma}')) = p((\sigma'', \sigma'''), (\sigma, \sigma'))$. This result again follows from the fact that taking the complement of a sequence essentially amounts to a relabelling of the bases, and a change in the direction that the sequence is read. Thus, all Hamming distances in Eq. (13) are unchanged.

Therefore, with this choice of landscape, and with a genome-independent $\epsilon$ and $\lambda$, we have, assuming that our quasispecies population is initially lesion-free (which is done by taking $y_{(\sigma_0, \bar{\sigma}_0)} = y_{(\bar{\sigma}_0, \sigma_0)} = 1/2$, for instance) that $y_{(\bar{\sigma}, \bar{\sigma}')} = y_{(\sigma, \sigma')}$, and so Eq. (9) applies.

In the limit of infinite sequence length, we claim that we may treat the quasispecies dynamics about the "master" pairs $(\sigma_0, \bar{\sigma}_0)$ and $(\bar{\sigma}_0, \sigma_0)$ independently of one another. That is, we may treat the dynamics as arising from essentially two separate quasispecies living on separate fitness landscapes. The heuristic reason for this is as follows: Because $\sigma_0$ and $\bar{\sigma}_0$ become infinitely separated from each other in the limit of infinite sequence length, it follows that if a given $\sigma''$ is of finite Hamming distance to either $\sigma_0$ (or $\bar{\sigma}_0$), then after daughter strand synthesis and lesion repair we obtain a $(\sigma, \sigma')$ which is of finite Hamming distance to $(\sigma_0, \bar{\sigma}_0)$ (or $(\bar{\sigma}_0, \sigma_0)$). This is because the probability of making an infinite number of replication mistakes is zero, and so if $\sigma''$ is of finite Hamming distance to $\sigma_0$, then it remains so after replication, and its complement is also of finite Hamming distance to $\bar{\sigma}_0$.

We allow our system to come to equilibrium from the initial condition $y_{(\sigma_0, \bar{\sigma}_0)} = y_{(\bar{\sigma}_0, \sigma_0)} = 1/2$ (equivalent to $x_{\{\sigma_0, \bar{\sigma}_0\}} = 1$. We choose this initial condition because it guarantees convergence to the unique stable equilibrium solution of the model. The reason for this is that all genomes are mutationally accessible from $\{\sigma_0, \bar{\sigma}_0\}$. Because of the neglect of backmutations in the limit of infinite sequence length, other initial conditions may lead to different regions of the genome space becoming mutationally disconnected from each other, preventing proper equilibration from occurring).

Thus, because our genome distribution is initially localized about $(\sigma_0, \bar{\sigma}_0)$ and $(\bar{\sigma}_0, \sigma_0)$, we need only consider the local dynamics about each master sequence pair, and treat the dynamics about each pair separately.

We now claim that, for sequence pairs $(\sigma, \sigma')$ of the first class, $y_{(\sigma, \sigma')}$ depends only on $l_C, l_L, l_R,$ and $l_B$. We note that this certainly holds at $t = 0$, given our initial conditions. In order to prove that this holds at all times, we need to show that, if $y_{(\sigma, \sigma')}$ depends only on $l_C, l_L, l_R,$ and $l_B$ at some time $t$, then $dy_{(\sigma, \sigma')}/dt$ depends only on $l_C, l_L, l_R,$ and $l_B$. In doing so, we will be simultaneously deriving the dynamical form of the quasispecies equations appropriate for our choice of fitness landscapes. We should note that our "proof" will not be strictly rigorous, since it will consider finite sequence length equations while still assuming that the first class and second class sequence pair dynamics may be treated as separate quasispecies. Nevertheless, since we are passing to the limit $L \to \infty$, we can assume that $L$ is sufficiently large to make the correction terms to our equations negligible, and eventually 0, in the limit.

So, suppose that at some time $t$, for all sequence pairs $(\sigma, \sigma')$ we have that $y_{(\sigma,\sigma')}$ depends only on $l_C$, $l_L$, $l_R$, and $l_B$. Then we may write $y_{(\sigma,\sigma')} = y_{(l_C,l_L,l_R,l_B)}$, and so, Eq. (9) gives,

$$\begin{aligned}\frac{dy_{(l_C,l_L,l_R,l_B)}}{dt} &= -(\kappa_{(l_C,l_L,l_R,l_B)} + \bar{\kappa}(t))y_{(l_C,l_L,l_R,l_B)} \\ &+ \sum_{(\sigma'',\sigma''')} \kappa_{(\sigma'',\sigma''')} y_{(\sigma'',\sigma''')} \times \\ &\quad p((\sigma'',\sigma'''),(\sigma,\sigma')) \\ &+ \sum_{(\sigma'',\sigma''')} \kappa_{(\sigma'',\sigma''')} y_{(\sigma'',\sigma''')} \times \\ &\quad p((\sigma'',\sigma'''),(\bar{\sigma}',\bar{\sigma}))\end{aligned} \quad (21)$$

We proceed as follows: Given $\sigma_C''$ and $\sigma_C$, then among the subset of positions where $\sigma_C$ and $\sigma_0$ are identical, let $l_{C,1}$ denote where $\sigma_C''$ differs from $\sigma_C$. Among the subset of positions where $\sigma_C$ and $\sigma_0$ differ, let $l_{C,2}$ denote where $\sigma_C''$ is identical to $\sigma_0$. Finally, where $\sigma_C$ differs from $\sigma_0$, let $l_{C,3}$ denote the number of positions where $\sigma_C''$ differs from both $\sigma_0$ and $\sigma_C$. It is clear that $D_H(\sigma_C'', \sigma_C) = l_{C,1} + l_{C,2} + l_{C,3}$. Furthermore, to have a nonzero value of $p((\sigma'',\sigma'''),(\sigma,\sigma'))$, we must have $\sigma_{NC}'' = \sigma_{NC}$. Since the sequence pair $(\sigma, \sigma')$ consists of $l_L + l_R + l_B$ lesions, it follows that $L_{NC} = l_L + l_R + l_B$, giving,

$$\begin{aligned}p((\sigma'',\sigma'''),(\sigma,\sigma')) &= \left(\frac{\lambda\epsilon}{2(S-1)}\right)^{l_{C,1}+l_{C,2}+l_{C,3}} \times \\ &\quad \left(1 - \epsilon\left(1-\frac{\lambda}{2}\right)\right)^{L-l_L-l_R-l_B-l_{C,1}-l_{C,2}-l_{C,3}} \times \\ &\quad \left(\frac{\epsilon(1-\lambda)}{S-1}\right)^{l_L+l_R+l_B}\end{aligned} \quad (22)$$

We now need to characterize the $\sigma'''$: Where $\sigma''$ differs from $\sigma_0$, let $l_1'''$ denote the number of sites where $\sigma''$ and $\sigma'''$ are complementary, and $l_2'''$ the number of sites where $\sigma'''$ is non-complementary to $\sigma''$ but differs from $\bar{\sigma}_0$. Let $l_3'''$ denote the number of sites where $\sigma''$ is identical to $\sigma_0$, where $\sigma'''$ is non-complementary to $\sigma''$. Then we have, $l_C''' = l_1'''$, $l_L''' = l_{C,1} + l_C + l_L + l_B - l_{C,2} - l_1''' - l_2'''$, $l_R''' = l_3'''$, and $l_B''' = l_2'''$.

We define $C''(l_{C,1}, l_{C,2}, l_{C,3}; l_C, l_L, l_R, l_B)$ to be the number of $\sigma''$ characterized by $l_{C,1}$, $l_{C,2}$, and $l_{C,3}$. We have,

$$C''(l_{C,1}, l_{C,2}, l_{C,3}; l_C, l_L, l_R, l_B) = \binom{L - l_L - l_R - l_B - l_C}{l_{C,1}} \binom{l_C}{l_{C,2}} \binom{l_C - l_{C,2}}{l_{C,3}} (S-1)^{l_{C,1}} (S-2)^{l_{C,3}} \quad (23)$$

where $l_{C,1}$ ranges from 0 to $L - l_L - l_R - l_B - l_C$, $l_{C,2}$ ranges from 0 to $l_C$, and $l_{C,3}$ ranges from 0 to $l_C - l_{C,2}$.

For each such choice of $\sigma''$, we define $C'''(l_1''', l_2''', l_3'''; l_C, l_L, l_R, l_B, l_{C,1}, l_{C,2}, l_{C,3})$ to be the number of $\sigma'''$ characterized by $l_1'''$, $l_2'''$, and $l_3'''$. We have,

$$\begin{aligned}C'''(l_1''', l_2''', l_3'''; l_C, l_L, l_R, l_B, l_{C,1}, l_{C,2}, l_{C,3}) &= \binom{l_{C,1} + l_L + l_B + l_C - l_{C,2}}{l_1'''} \binom{l_{C,1} + l_L + l_B + l_C - l_{C,2} - l_1'''}{l_2'''} \times \\ &\quad \binom{L - l_{C,1} - l_L - l_B - l_C + l_{C,2}}{l_3'''} (S-2)^{l_2'''} (S-1)^{l_3'''}\end{aligned} \quad (24)$$

We may perform a similar analysis on $(\bar{\sigma}', \bar{\sigma})$, which is characterized by the parameters $l_C$, $l_R$, $l_L$, and $l_B$. The quasispecies equations then become,

$$\begin{aligned}
\frac{dy_{(l_C,l_L,l_R,l_B)}}{dt} &= -(\kappa_{(l_C,l_L,l_R,l_B)} + \bar{\kappa}(t))y_{(l_C,l_L,l_R,l_B)} \\
&+ \sum_{l_{C,1}=0}^{L-l_L-l_R-l_B-l_C} \sum_{l_{C,2}=0}^{l_C} \sum_{l_{C,3}=0}^{l_C-l_{C,2}} \sum_{l''_1=0}^{l_{C,1}+l_L+l_B+l_C-l_{C,2}} \sum_{l''_2=0}^{l_{C,1}+l_L+l_B+l_C-l_{C,2}-l''_1} \sum_{l''_3=0}^{L-l_{C,1}-l_L-l_B-l_C+l_{C,2}} \times \\
&\quad C'''(l_{C,1},l_{C,2},l_{C,3};l_C,l_L,l_R,l_B) C''''(l''_1,l''_2,l''_3;l_C,l_L,l_R,l_B,l_{C,1},l_{C,2},l_{C,3}) \times \\
&\quad \kappa_{(l''_1,l_{C,1}+l_L+l_B+l_C-l_{C,2}-l''_1-l''_2,l''_3,l''_2)} y_{(l''_1,l_{C,1}+l_L+l_B+l_C-l_{C,2}-l''_1-l''_2,l''_3,l''_2)} \times \\
&\quad (\frac{\lambda\epsilon}{2(S-1)})^{l_{C,1}+l_{C,2}+l_{C,3}} (\frac{\epsilon(1-\lambda)}{S-1})^{l_L+l_R+l_B} (1-\epsilon(1-\frac{\lambda}{2}))^{L-l_L-l_R-l_B-l_{C,1}-l_{C,2}-l_{C,3}} \\
&+ \sum_{l_{C,1}=0}^{L-l_L-l_R-l_B-l_C} \sum_{l_{C,2}=0}^{l_C} \sum_{l_{C,3}=0}^{l_C-l_{C,2}} \sum_{l''_1=0}^{l_{C,1}+l_R+l_B+l_C-l_{C,2}} \sum_{l''_2=0}^{l_{C,1}+l_R+l_B+l_C-l_{C,2}-l''_1} \sum_{l''_3=0}^{L-l_{C,1}-l_R-l_B-l_C+l_{C,2}} \times \\
&\quad C'''(l_{C,1},l_{C,2},l_{C,3};l_C,l_R,l_L,l_B) C''''(l''_1,l''_2,l''_3;l_C,l_R,l_L,l_B,l_{C,1},l_{C,2},l_{C,3}) \times \\
&\quad \kappa_{(l''_1,l_{C,1}+l_R+l_B+l_C-l_{C,2}-l''_1-l''_2,l''_3,l''_2)} y_{(l''_1,l_{C,1}+l_R+l_B+l_C-l_{C,2}-l''_1-l''_2,l''_3,l''_2)} \times \\
&\quad (\frac{\lambda\epsilon}{2(S-1)})^{l_{C,1}+l_{C,2}+l_{C,3}} (\frac{\epsilon(1-\lambda)}{S-1})^{l_L+l_R+l_B} (1-\epsilon(1-\frac{\lambda}{2}))^{L-l_L-l_R-l_B-l_{C,1}-l_{C,2}-l_{C,3}} \quad (25)
\end{aligned}$$

We now use the binomial theorem and sum over $l_{C,3}$, giving,

$$\begin{aligned}
\frac{dy_{(l_C,l_L,l_R,l_B)}}{dt} &= -(\kappa_{(l_C,l_L,l_R,l_B)} + \bar{\kappa}(t))y_{(l_C,l_L,l_R,l_B)} \\
&+ \sum_{l_{C,1}=0}^{L-l_L-l_R-l_B-l_C} \sum_{l_{C,2}=0}^{l_C} \sum_{l''_1=0}^{l_{C,1}+l_L+l_B+l_C-l_{C,2}} \sum_{l''_2=0}^{l_{C,1}+l_L+l_B+l_C-l_{C,2}-l''_1} \sum_{l''_3=0}^{L-l_{C,1}-l_L-l_B-l_C+l_{C,2}} \times \\
&\quad \binom{L-l_L-l_R-l_B-l_C}{l_{C,1}} \binom{l_C}{l_{C,2}} \binom{l_{C,1}+l_L+l_B+l_C-l_{C,2}}{l''_1} \binom{l_{C,1}+l_L+l_B+l_C-l_{C,2}-l''_1}{l''_2} \times \\
&\quad \binom{L-l_{C,1}-l_L-l_B-l_C+l_{C,2}}{l''_3} (S-1)^{l''_3}(S-2)^{l''_2} \times \\
&\quad \kappa_{(l''_1,l_{C,1}+l_L+l_B+l_C-l_{C,2}-l''_1-l''_2,l''_3,l''_2)} y_{(l''_1,l_{C,1}+l_L+l_B+l_C-l_{C,2}-l''_1-l''_2,l''_3,l''_2)} \times \\
&\quad (\frac{\lambda\epsilon}{2})^{l_{C,1}} (\frac{\lambda\epsilon}{2(S-1)})^{l_{C,2}} (\frac{\epsilon(1-\lambda)}{S-1})^{l_L+l_R+l_B} \times \\
&\quad (1-\epsilon(1-\frac{\lambda}{2}))^{L-l_L-l_R-l_B-l_C-l_{C,1}} (1-\epsilon(1-\lambda(1-\frac{1}{2(S-1)})))^{l_C-l_{C,2}} \\
&+ \sum_{l_{C,1}=0}^{L-l_L-l_R-l_B-l_C} \sum_{l_{C,2}=0}^{l_C} \sum_{l''_1=0}^{l_{C,1}+l_R+l_B+l_C-l_{C,2}} \sum_{l''_2=0}^{l_{C,1}+l_R+l_B+l_C-l_{C,2}-l''_1} \sum_{l''_3=0}^{L-l_{C,1}-l_R-l_B-l_C+l_{C,2}} \times \\
&\quad \binom{L-l_L-l_R-l_B-l_C}{l_{C,1}} \binom{l_C}{l_{C,2}} \binom{l_{C,1}+l_R+l_B+l_C-l_{C,2}}{l''_1} \binom{l_{C,1}+l_R+l_B+l_C-l_{C,2}-l''_1}{l''_2} \times \\
&\quad \binom{L-l_{C,1}-l_R-l_B-l_C+l_{C,2}}{l''_3} (S-1)^{l''_3}(S-2)^{l''_2} \times \\
&\quad \kappa_{(l''_1,l_{C,1}+l_R+l_B+l_C-l_{C,2}-l''_1-l''_2,l''_3,l''_2)} y_{(l''_1,l_{C,1}+l_R+l_B+l_C-l_{C,2}-l''_1-l''_2,l''_3,l''_2)} \times \\
&\quad (\frac{\lambda\epsilon}{2})^{l_{C,1}} (\frac{\lambda\epsilon}{2(S-1)})^{l_{C,2}} (\frac{\epsilon(1-\lambda)}{S-1})^{l_L+l_R+l_B} \times \\
&\quad (1-\epsilon(1-\frac{\lambda}{2}))^{L-l_L-l_R-l_B-l_C-l_{C,1}} (1-\epsilon(1-\lambda(1-\frac{1}{2(S-1)})))^{l_C-l_{C,2}} \quad (26)
\end{aligned}$$

Note that this expression depends only on $l_C$, $l_L$, $l_R$, and $l_B$, hence our claim that $y_{(\sigma,\sigma')}$ only depends on $l_C$, $l_L$, $l_R$, and $l_B$ is established. We now proceed to formally take the $L \to \infty$ limit.

Because $y_{(\sigma,\sigma')}$ depends only on $l_C$, $l_L$, $l_R$, and $l_B$, we can sum over the population fractions of all first class sequence pairs characterized by a given set of $l_C$, $l_L$, $l_R$, and $l_B$, and reexpress the quasispecies dynamics in terms of these quantities. To do this, we define $C(l_C, l_L, l_R, l_B)$ to be the number of sequence pairs characterized by $l_C, l_L, l_R, l_B$, and note that,

$$C(l_C, l_L, l_R, l_B) = \frac{L!}{l_C! l_L! l_R! l_B! (L - l_C - l_L - l_R - l_B)!} (S-1)^{l_C + l_L + l_R + l_B} (S-2)^{l_B} \qquad (27)$$

We let $z_{(l_C, l_L, l_R, l_B)}$ denote the total population fraction of first class sequence pairs characterized by $l_C, l_L, l_R$, and $l_B$, so that $z_{(l_C, l_L, l_R, l_B)} = C(l_C, l_L, l_R, l_B) y_{(l_C, l_L, l_R, l_B)}$. From this expression, and using the fact that $C_{(l_C, l_R, l_L, l_B)} = C_{(l_C, l_L, l_R, l_B)}$, we may reexpress the quasispecies dynamics in terms of the $z_{(l_C, l_L, l_R, l_B)}$. After some algebra, the final result is,

$$\begin{aligned}
\frac{dz_{(l_C, l_L, l_R, l_B)}}{dt} =& -(\kappa_{(l_C, l_L, l_R, l_B)} + \bar{\kappa}(t)) z_{(l_C, l_L, l_R, l_B)} \\
& + \sum_{l_{C,1}=0}^{L-l_L-l_R-l_B-l_C} \sum_{l_{C,2}=0}^{l_C} \sum_{l_1''=0}^{l_{C,1}+l_L+l_B+l_C-l_{C,2}} \sum_{l_2''=0}^{l_{C,1}+l_L+l_B+l_C-l_{C,2}-l_1''} \sum_{l_3''=0}^{L-l_{C,1}-l_L-l_B-l_C+l_{C,2}} \times \\
& \binom{l_L + l_B + l_{C,1} + l_C - l_{C,2}}{l_{C,1}} \binom{l_L + l_B + l_C - l_{C,2}}{l_C - l_{C,2}} \binom{l_L + l_B}{l_L} \times \\
& \binom{L - l_L - l_B - l_{C,1} - l_C + l_{C,2}}{l_{C,2}} \binom{L - l_L - l_B - l_{C,1} - l_C}{l_R} \times \\
& \left(\frac{\lambda \epsilon}{2(S-1)}\right)^{l_{C,1}} \left(\frac{\lambda \epsilon}{2}\right)^{l_{C,2}} \left(1 - \epsilon\left(1 - \frac{\lambda}{2}\right)\right)^{L - l_L - l_R - l_B - l_C - l_{C,1}} \left(1 - \epsilon\left(1 - \lambda\left(1 - \frac{1}{2(S-1)}\right)\right)\right)^{l_C - l_{C,2}} \times \\
& (\epsilon(1-\lambda))^{l_R} \left(\frac{\epsilon(1-\lambda)}{S-1}\right)^{l_L} \left(\frac{\epsilon(1-\lambda)(S-2)}{S-1}\right)^{l_B} \times \\
& \kappa_{(l_1'', l_{C,1}+l_L+l_B+l_C-l_{C,2}-l_1''-l_2'', l_3'', l_2'')} z_{(l_1'', l_{C,1}+l_L+l_B+l_C-l_{C,2}-l_1''-l_2'', l_3'', l_2'')} \\
& + \sum_{l_{C,1}=0}^{L-l_L-l_R-l_B-l_C} \sum_{l_{C,2}=0}^{l_C} \sum_{l_1''=0}^{l_{C,1}+l_R+l_B+l_C-l_{C,2}} \sum_{l_2''=0}^{l_{C,1}+l_R+l_B+l_C-l_{C,2}-l_1''} \sum_{l_3''=0}^{L-l_{C,1}-l_R-l_B-l_C+l_{C,2}} \times \\
& \binom{l_R + l_B + l_{C,1} + l_C - l_{C,2}}{l_{C,1}} \binom{l_R + l_B + l_C - l_{C,2}}{l_C - l_{C,2}} \binom{l_R + l_B}{l_R} \times \\
& \binom{L - l_R - l_B - l_{C,1} - l_C + l_{C,2}}{l_{C,2}} \binom{L - l_R - l_B - l_{C,1} - l_C}{l_L} \times \\
& \left(\frac{\lambda \epsilon}{2(S-1)}\right)^{l_{C,1}} \left(\frac{\lambda \epsilon}{2}\right)^{l_{C,2}} \left(1 - \epsilon\left(1 - \frac{\lambda}{2}\right)\right)^{L - l_L - l_R - l_B - l_C - l_{C,1}} \left(1 - \epsilon\left(1 - \lambda\left(1 - \frac{1}{2(S-1)}\right)\right)\right)^{l_C - l_{C,2}} \times \\
& (\epsilon(1-\lambda))^{l_L} \left(\frac{\epsilon(1-\lambda)}{S-1}\right)^{l_R} \left(\frac{\epsilon(1-\lambda)(S-2)}{S-1}\right)^{l_B} \times \\
& \kappa_{(l_1'', l_{C,1}+l_R+l_B+l_C-l_{C,2}-l_1''-l_2'', l_3'', l_2'')} z_{(l_1'', l_{C,1}+l_R+l_B+l_C-l_{C,2}-l_1''-l_2'', l_3'', l_2'')}
\end{aligned} \qquad (28)$$

Now, it may be shown in the limit of infinite sequence length that only the $l_{C,1} = 0$ terms contribute to the sum. This corresponds to the neglect of backmutations in the limit of infinite sequence length. The proof that $l_{C,1} > 0$ terms may be neglected is fairly tedious, but is similar to the arguments given in [25, 27]. Therefore, we do not give details in this paper. Regarding the remaining terms, we may note that,

$$\begin{aligned}
\binom{L - l_L - l_B - l_C + l_{C,2}}{l_{C,2}} \left(\frac{\lambda \epsilon}{2}\right)^{l_{C,2}} \left(1 - \epsilon\left(1 - \frac{\lambda}{2}\right)\right)^{L - l_L - l_R - l_B - l_C} &\to \frac{1}{l_{C,2}!} \left(\frac{\lambda \mu}{2}\right)^{l_{C,2}} e^{-\mu(1-\frac{\lambda}{2})} \\
\binom{L - l_L - l_B - l_C}{l_R} (\epsilon(1-\lambda))^{l_R} &\to \frac{1}{l_R!} (\mu(1-\lambda))^{l_R} \\
\left(1 - \epsilon\left(1 - \lambda\left(1 - \frac{1}{2(S-1)}\right)\right)\right)^{l_C - l_{C,2}} &\to 1 \\
\binom{l_L + l_B + l_C - l_{C,2}}{l_C - l_{C,2}} \binom{l_L + l_B}{l_L} \left(\frac{\epsilon(1-\lambda)(S-2)}{S-1}\right)^{l_B} \left(\frac{\epsilon(1-\lambda)}{S-1}\right)^{l_L} &\to \delta_{l_L, 0} \delta_{l_B, 0}
\end{aligned} \qquad (29)$$

The last statement implies that genomes with $l_B > 0$ and genomes with $l_L, l_R$ simultaneously $> 0$ cannot be produced by replication. Therefore, if our initial population distribution is such that $z_{(l_C, l_L, l_R, l_B > 0)} = 0$ and $z_{(l_C, l_L>0, l_R>0, l_B)} = 0$ (as is the case with our initial conditions), then we may assume that $z_{(l_C, l_L, l_R, l_B>0)} = 0$ and $z_{(l_C, l_L>0, l_R>0, l_B)} = 0$ at all times.

Putting everything together, we obtain the final form of the infinite sequence length equations,

$$\begin{aligned}
\frac{dz_{(l_C, 0, 0, 0)}}{dt} &= -(\kappa_{(l_C, 0, 0, 0)} + \bar{\kappa}(t)) z_{(l_C, 0, 0, 0)} \\
&\quad + 2 e^{-\mu(1-\frac{\lambda}{2})} \sum_{l'_C=0}^{l_C} \frac{1}{l'_C!} (\frac{\lambda\mu}{2})^{l'_C} \sum_{l''_1=0}^{l_C-l'_C} \sum_{l''_2=0}^{\infty} \times \\
&\quad \kappa_{(l''_1, l_C-l'_C-l''_1, l''_2, 0)} z_{(l''_1, l_C-l'_C-l''_1, l''_2, 0)} \\
\frac{dz_{(l_C, l_L>0, 0, 0)}}{dt} &= -(\kappa_{(l_C, l_L, 0, 0)} + \bar{\kappa}(t)) z_{(l_C, l_L, 0, 0)} \\
&\quad + \frac{1}{l_L!} (\mu(1-\lambda))^{l_L} e^{-\mu(1-\frac{\lambda}{2})} \sum_{l'_C=0}^{l_C} \frac{1}{l'_C!} (\frac{\lambda\mu}{2})^{l'_C} \sum_{l''_1=0}^{l_C-l'_C} \sum_{l''_2=0}^{\infty} \kappa_{(l''_1, l_C-l'_C-l''_1, l''_2, 0)} z_{(l''_1, l_C-l'_C-l''_1, l''_2, 0)} \\
\frac{dz_{(l_C, 0, l_R>0, 0)}}{dt} &= -(\kappa_{(l_C, 0, l_R, 0)} + \bar{\kappa}(t)) z_{(l_C, 0, l_R, 0)} \\
&\quad + \frac{1}{l_R!} (\mu(1-\lambda))^{l_R} e^{-\mu(1-\frac{\lambda}{2})} \sum_{l'_C=0}^{l_C} \frac{1}{l'_C!} (\frac{\lambda\mu}{2})^{l'_C} \sum_{l''_1=0}^{l_C-l'_C} \sum_{l''_2=0}^{\infty} \kappa_{(l''_1, l_C-l'_C-l''_1, l''_2, 0)} z_{(l''_1, l_C-l'_C-l''_1, l''_2, 0)}
\end{aligned} \quad (30)$$

The reason why genomes where both $l_L$ and $l_R$ are nonzero, or where $l_B$ is nonzero, cannot be produced by replication, is as follows: Given a parent strand $\sigma$ which differs from $\sigma_0$ in $l$ places, the probability of correct daughter strand synthesis in these $l$ places is $(1-\epsilon)^l = (1-\mu/L)^l \to 1$ as $L \to \infty$. Therefore, any mismatches that occur will occur where $\sigma$ and $\sigma_0$ are identical. Wherever lesion repair does not occur, $\sigma$ remains identical to $\sigma_0$ in the final genome. The result is a sequence pair for which $l_L = l_B = 0$. Similarly, a parent strand $\sigma$ which differs from $\bar{\sigma}_0$ in a finite number of positions produces a sequence pair for which $l_R = l_B = 0$. Therefore, as is reflected in the equations, it is impossible for replication to produce sequence pairs for which $l_L$ and $l_R$ are simultaenously nonzero, or for which $l_B$ is nonzero. Since the population fractions of these genomes is initially 0, they remain 0 for all time, hence we may simply assume that $z_{(l_C, l_L, l_R, l_B)} = 0$ if $l_L$ and $l_R$ are simultaneously nonzero, or if $l_B$ is nonzero.

These equations describe the quasispecies dynamics for the first-class sequence pairs. An analogous set of equations may be derived for the second-class sequence pairs, where we let $\bar{z}_{(l_C, l_L, l_R, l_B)}$ denote the total population fraction of second-class sequence pairs characterized by the parameters $l_C$, $l_L$, $l_R$, and $l_B$. Note that a sequence pair $(\sigma, \sigma')$ is of the second class if and only if $(\sigma', \sigma)$ is of the first class. Therefore, since $y_{(\sigma, \sigma')} = y_{(\sigma', \sigma)}$, it follows that $\bar{z}_{(l_C, l_L, l_R, l_B)} = z_{(l_C, l_R, l_L, l_B)}$.

We can provide an expression for the mean fitness in terms of the $z_{(l_C, l_L, l_R, l_B)}$. First note that, since the total population fraction of the third class sequence pairs is $1 - \sum_{l_C=0}^{\infty} \sum_{l_L=0}^{\infty} \sum_{l_R=0}^{\infty} \sum_{l_B=0}^{\infty} (z_{(l_C, l_L, l_R, l_B)} + \bar{z}_{(l_C, l_L, l_R, l_B)})$, it follows that the mean fitness is given by,

$$\begin{aligned}
\bar{\kappa}(t) &= \sum_{l_C=0}^{\infty} \sum_{l_L=0}^{\infty} \sum_{l_R=0}^{\infty} \sum_{l_B=0}^{\infty} \times \\
&\quad (\kappa_{(l_C, l_L, l_R, l_B)} z_{(l_C, l_L, l_R, l_B)} + \\
&\quad \kappa_{(l_C, l_R, l_L, l_B)} \bar{z}_{(l_C, l_L, l_R, l_B)}) + \\
&\quad 1 - \sum_{l_C=0}^{\infty} \sum_{l_L=0}^{\infty} \sum_{l_R=0}^{\infty} \sum_{l_B=0}^{\infty} \times \\
&\quad (z_{(l_C, l_L, l_R, l_B)} + \bar{z}_{(l_C, l_L, l_R, l_B)})
\end{aligned} \quad (31)$$

For our particular class of fitness landscapes, for which we have $y_{(\bar{\sigma}, \bar{\sigma}')} = y_{(\sigma, \sigma')}$, we get $y_{(l_C, l_L, l_R, l_B)} = y_{(\bar{\sigma}', \bar{\sigma})} = y_{(\sigma, \sigma')} = y_{(l_C, l_L, l_R, l_B)}$, and so $\bar{z}_{(l_C, l_L, l_R, l_B)} = z_{(l_C, l_L, l_R, l_B)}$. This allows us to reexpress the expression for the mean fitness as,

$$\bar{\kappa}(t) = 2 \sum_{l_C=0}^{\infty} \sum_{l_L=0}^{\infty} \sum_{l_R=0}^{\infty} \sum_{l_B=0}^{\infty} (\kappa_{(l_C, l_L, l_R, l_B)} - 1) z_{(l_C, l_L, l_R, l_B)} + 1 \quad (32)$$

## IV. SOLUTION OF THE GENERALIZED SINGLE-FITNESS-PEAK LANDSCAPE

### A. Equilibrium mean fitness and the error catastrophe

The simplest and most commonly studied landscape in the quasispecies model is known as the single-fitness-peak landscape. For the single-stranded RNA genomes modeled in the original quasispecies equations, this landscape is defined by a "master" genome $\sigma_0$ with a first-order growth rate constant $k > 1$, while all other genomes have a first-order growth rate constant of 1. Thus, the master genome is said to be viable, while all the other genomes are unviable.

For semiconservatively replicating, double-stranded DNA genomes, the single fitness peak landscape is defined by a "master" genome $\{\sigma_0, \bar{\sigma}_0\}$. When converting the quasispecies dynamics from the space of genomes to the space of single strands, the resulting single fitness peak landscape becomes a two-peak landscape with "master" sequences $\sigma_0$ and $\bar{\sigma}_0$.

For imperfect lesion repair, it is therefore also natural for us to first study the single-fitness-peak landscape. In this section, instead of considering a single-fitness-peak landscape where any change to the "master" genome $\{\sigma_0, \bar{\sigma}_0\}$ results in an unviable genome, we consider a "generalized" single fitness peak landscape, where the master genome can sustain a finite number of lesions and remain viable.

In the limit of infinite sequence length, the $l$-lesion landscape may therefore be defined as follows: For sequence pairs of the first class, we define $\kappa_{(l_C, l_L, l_R, l_B)} = k > 1$ if $l_C = 0$ and $l_L + l_R + l_B \leq l$, otherwise $\kappa_{(l_C, l_L, l_R, l_B)} = 1$. The landscape of sequence pairs of the second class is of course defined by the landscape for sequence pairs of the first class, via $\kappa_{(\sigma, \sigma')} = \kappa_{(\sigma', \sigma)}$. All sequence pairs of the third class are unviable.

Because we may make the assumption that $z_{(l_C, l_L, l_R, l_B)} = 0$ if $l_B \neq 0$ or if both $l_L$ and $l_R$ are nonzero, we need only consider $z_{(l_C, l_L, 0, 0)}$ and $z_{(l_C, 0, l_R, 0)}$. Furthermore, by the symmetry of our landscape we have $z_{(l_C, l', 0, 0)} = z_{(l_C, 0, l', 0)}$.

We define the following quantities for use in our calculations:

$$z_0 = z_{(0,0,0,0)} + \sum_{l_L=1}^{l} z_{(0,l_L,0,0)} + \sum_{l_R=1}^{l} z_{(0,0,l_R,0)}$$
$$= z_{(0,0,0,0)} + 2 \sum_{l'=1}^{l} z_{(0,0,l',0)} \quad (33)$$

$$z_1 = \sum_{l'=0}^{l} z_{(0,0,l',0)} \quad (34)$$

$$z_2 = \sum_{l'=0}^{\infty} z_{(0,0,l',0)} \quad (35)$$

We then have, from Eq. (30), that,

$$\begin{aligned}
\frac{dz_{(0,0,0,0)}}{dt} &= -(k + \bar{\kappa}(t))z_{(0,0,0,0)} \\
&\quad + 2e^{-\mu(1-\frac{\lambda}{2})}[(k-1)z_1 + z_2] \\
\frac{dz_1}{dt} &= -(k + \bar{\kappa}(t))z_1 \\
&\quad + e^{-\mu(1-\frac{\lambda}{2})}(1 + f_l(\mu, \lambda))[(k-1)z_1 + z_2] \\
\frac{dz_2}{dt} &= -\bar{\kappa}(t)z_2 \\
&\quad + (e^{-\mu\frac{\lambda}{2}} + e^{-\mu(1-\frac{\lambda}{2})} - 1)[(k-1)z_1 + z_2]
\end{aligned} \quad (36)$$

where $f_l(\mu, \lambda) \equiv \sum_{l'=0}^{l} \frac{1}{l'!}[\mu(1-\lambda)]^{l'}$.

Now, note that $z_0 = z_{(0,0,0,0)} + 2(z_1 - z_{(0,0,0,0)})$. Furthermore, note from Eq. (32) that $\bar{\kappa}(t) = k(2z_0) + (1 - 2z_0) = 2(k-1)z_0 + 1$. Setting the left-hand side of Eq. (36) to 0, we may systematically eliminate variables to obtain,

$$0 = -\frac{z_1}{\bar{\kappa}(t=\infty) - (e^{-\mu\frac{\lambda}{2}} + e^{-\mu(1-\frac{\lambda}{2})} - 1)} \times$$
$$(\bar{\kappa}(t=\infty)^2 - A(\mu, \lambda)\bar{\kappa}(t=\infty) - B(\mu, \lambda)) \quad (37)$$

where,

$$\begin{aligned}
A(\mu, \lambda) &= k((1 + f_l(\mu, \lambda))e^{-\mu(1-\frac{\lambda}{2})} - 1) \\
&\quad - f_l(\mu, \lambda)e^{-\mu(1-\frac{\lambda}{2})} + e^{-\mu\frac{\lambda}{2}} - 1 \\
B(\mu, \lambda) &= k(e^{-\mu\frac{\lambda}{2}} + e^{-\mu(1-\frac{\lambda}{2})} - 1) \quad (38)
\end{aligned}$$

Eq. (37) admits multiple solutions. To determine the physical solution at a given $\mu$, we note that we want $\bar{\kappa}(t=\infty) = k$ for $\mu = 0$. This simply reflects the fact that when replication is error-free, the population consists entirely of viable genomes. Therefore, for sufficiently small $\mu$, the equilibrium mean fitness is given by,

$$\bar{\kappa}(t=\infty) = \frac{A(\mu,\lambda) + \sqrt{A(\mu,\lambda)^2 + 4B(\mu,\lambda)}}{2} \quad (39)$$

The equilibrium mean fitness is given by this expression until the error catastrophe, which occurs when the value of $\bar{\kappa}(t=\infty)$ given by the formula above equals 1. At this point, the selective advantage for remaining viable is no longer sufficiently strong to localize the population about the viable genomes. The fraction of viable genomes drops to 0, and the fitness of the population simply becomes the fitness of the unviable genomes.

Setting $\bar{\kappa}(t=\infty) = 1$ in Eq. (37), it is possible to show, after some manipulation, that the critical value of $\mu$, denoted $\mu_{crit}$, is the solution to the equation,

$$\frac{e^{-\mu(1-\frac{\lambda}{2})}}{2 - e^{-\mu\frac{\lambda}{2}}} = \frac{k+1}{k(2 + f_l(\mu, \lambda)) - f_l(\mu, \lambda)} \quad (40)$$

## V. RESULTS AND DISCUSSION

### A. Behavior of the model for specific values of $l$ and $\lambda$

#### 1. $l = 0$

When $l = 0$, our fitness landscape corresponds to a single fitness peak landscape which tolerates no lesions. We have $f_0(\mu, \lambda) = 1$, so, as $k \to \infty$, we have, at the error catastrophe, that

$$\frac{e^{-\mu(1-\frac{\lambda}{2})}}{2 - e^{-\mu\frac{\lambda}{2}}} = \frac{1}{3} \quad (41)$$

When $\lambda = 0$, we obtain $e^{-\mu_{crit}} = 1/3$, while when $\lambda = 1$, we obtain $e^{-\mu_{crit}/2} = 1/2$. Note that, after daughter strand synthesis and lesion repair, the probability that a given parent base is matched up with the proper daughter base is given by $1 - \epsilon + (\lambda/2)\epsilon$. The reason for this is that correct base pair synthesis occurs with probability $1 - \epsilon$. A mismatch occurs with probability $\epsilon$, which is correctly repaired during lesion repair with probability $\lambda/2$. In the limit of infinite sequence length, the probability of correct daughter strand synthesis then becomes $\lim_{L \to \infty} (1 - \epsilon(1 - \lambda/2))^L = e^{-\mu(1-\lambda/2)}$.

Note then that for $\lambda = 0$, the critical daughter strand synthesis probability is lower than for $\lambda = 1$. The reason for this is that when lesion repair is turned off, the parent strands are unaffected by the replication process, and hence the information in the master genome is preserved by replication. Thus, although viability may be lost through erroneous replication, preserving the information in the parent strand makes it possible to recover a viable genome in the next replication cycle. The result is a delay in the critical replication fidelity to a lower value of $e^{-\mu_{crit}}$ than would be expected if it were assumed that unviable genomes cannot replicate into viable ones (the expected value from such an assumption is $1/2$).

#### 2. $l = \infty$

For $l = \infty$, our fitness landscape is one where only one of the "master" strands is necessary to confer viability. In this case, we have $f_\infty(\mu, \lambda) = e^{\mu(1-\lambda)}$, which gives $A(\mu, \lambda) = B(\mu, \lambda) - 1$. Below the error catastrophe, we therfore have,

$$\bar{\kappa}(t = \infty) = B(\mu, \lambda) = k(e^{-\mu\frac{\lambda}{2}} + e^{-\mu(1-\frac{\lambda}{2})} - 1) \quad (42)$$

Note that for $\lambda = 0$ we obtain $\bar{\kappa}(t = \infty) = ke^{-\mu}$. Thus, when lesion repair is turned off, we obtain an effectively conservatively replicating system. The error catastrophe in this case can be delayed to arbitrarily high mutation rates by increasing the replication rate of the viable genomes. This result was first derived by Brumer [30], and led to the hypothesis that imperfect lesion repair could reconcile semiconservative replication with the high mutation rates observed in many cancers (specifically the Microsatellite INstability, or MIN, tumors).

### B. Similarities to both conservative and semiconservative replication

Semiconservative replication with imperfect lesion repair bears a number of similarities and differences with semiconservative replication with perfect lesion repair and to conservative replication. We have shown earlier that the original semiconservative equations are obtained when $\lambda = 1$. Furthermore, we have also shown that when lesion repair is turned off, then if the fitness depends on only one of the strands, a semiconservatively replicating population becomes an effectively conservatively replicating one. For arbitrary lesion repair probabilities and for a given maximum lesion value $l$, it is interesting to explore what feature from both semiconservative and conservative replication are retained.

There are two key differences between conservative and semiconservative replication which we will explore here. First of all, for the single fitness peak landscape, the equilibrium mean fitness of a conservatively replicating system below the error catastrophe is $ke^{-\mu}$, which gives $\mu_{crit} = \ln k$. For a semiconservatively replicating system, we have an equilibrium mean fitness of $k(2e^{-\mu/2} - 1)$, which gives $\mu_{crit} = 2\ln 2/(1+1/k)$. Note that, as $k \to \infty$, $\mu_{crit} \to \infty$ for a conservative system, while for a semiconservative system, $\mu_{crit} \to 2\ln 2$. Thus, for a conservatively replicating system, the error threshold can be pushed to arbitrarily high mutation rates by making the growth rate of the master genome arbitrarily large. For semiconservative replication, in contrast, there is a maximal value to the error threshold. If the mutation rate exceeds this value, then no quasispecies will exist, independent of the growth rate constant of the viable genomes.

The reason for this difference in behavior is that conservative replication preserves a copy of the original genome. Therefore, no matter how high the mutation rate, by replicating fast enough, it is possible to produce viable genomes at a sufficient rate to out-replicate the unviable genomes, and thereby localize the population to a well-defined quasispecies. With semiconservative replication, the original genome is destroyed by the replication process. Therefore, on average, it is necessary for a viable genome to produce at least one viable copy per replication cycle. Otherwise, the net growth rate of the viable genomes becomes negative, and replicating faster simply kills off the viable population more quickly.

For arbitrary lesion repair probabilities, we can determine the value of $\mu_{crit}$ in the limit of $k \to \infty$ by solving,

$$\frac{e^{-\mu(1-\frac{\lambda}{2})}}{2 - e^{-\mu\frac{\lambda}{2}}} = \frac{1}{2 + f_l(\mu, \lambda)} \quad (43)$$

which may be rearranged to give,

$$(2 + f_l(\mu, \lambda))e^{-\mu(1-\frac{\lambda}{2})} + e^{-\mu\frac{\lambda}{2}} - 2 = 0 \quad (44)$$

When $\mu = 0$, $f_l(\mu, \lambda) = 1$, so the left-hand side evaluates to 2. When $\mu = \infty$, the left-hand side evaluates to $\lim_{\mu \to \infty} (f_l(\mu, \lambda)e^{-\mu(1-\frac{\lambda}{2})} + e^{-\mu\frac{\lambda}{2}} - 2)$. For *finite* $l$, $f_l(\mu, \lambda)$ is a polynomial, hence we obtain a limit of $-2$ for $\lambda > 0$, and a limit of $-1$ for $\lambda = 0$. Therefore, by the Intermediate Value Theorem, Eq. (44) has a solution, and so even as $k \to \infty$, $\mu_{crit}$ remains finite. This means that, for all finite $l$, semiconservative replication with arbitrary lesion repair is similar to the original semiconservative model in that there is an upper limit to the mutation rate before the error catastrophe occurs, independent of the growth rate of the viable genomes.

If $l = \infty$, then $f_l(\mu, \lambda) = e^{\mu(1-\lambda)}$, which gives $\lim_{\mu \to \infty} 2(e^{-\mu\frac{\lambda}{2}} - 1)$. When $\lambda > 0$, this limit is $-2$, so again $\mu_{crit}$ remains finite. When $\lambda = 0$, note that Eq. (44) evaluates to $2e^{-\mu} = 0$, which has no solution for finite $\mu$.

Therefore, unless $\lambda = 0$ and $l = \infty$, semiconservative replication with arbitrary lesion repair also has an upper bound to the mutation rate before the error catastrophe occurs. This makes sense, because, to ensure that after replication at least one of the daughter genomes is viable, it is necessary to prevent lesion repair from creating an unviable genome, and it is necessary to prevent lesions from destroying viability.

The second feature of semiconservative and conservative replication which we will consider has to do with the behavior of $\bar{\kappa}(t = \infty)$ near the error catastrophe. To make matters concrete, define $\kappa_{equil}(\mu) = \bar{\kappa}(t = \infty)$, and let us consider the behavior of $\kappa'_{equil}(\mu)$ for $\mu \to \mu^-_{crit}$.

For conservative replication, $\kappa'_{equil}(\mu) = -ke^{-\mu}$, so $\lim_{\mu \to \mu^-_{crit}} \kappa'_{equil}(\mu) = -1$. For semiconservative replication, $\kappa'_{equil}(\mu) = -ke^{-\mu/2}$, so $\lim_{\mu \to \mu^-_{crit}} \kappa'_{equil}(\mu) = -(k+1)/2$. As $k \to \infty$, this derivative goes to $-\infty$. In Appendix A, we will show that unless $\lambda = 1$, $\kappa'_{equil}(\mu)$ remains finite as $\mu \to \mu^-_{crit}$, assuming $l$ is finite.

In this sense, then, imperfect lesion repair is similar to conservative replication. The reason for this behavior is that, when lesion repair is imperfect, the correlation between the parent and daughter strands is broken. Therefore, it is possible that an erroneous daughter strand is synthesized, but that the errors are not communicated to the parent strand. On a subsequent replication cycle, the undamaged parent strand may be reintegrated into a master genome. Near the error catastrophe, where the efffective growth rate of the viable genomes is close to that of the unviable genomes, this effect slows down the rate at which the fitness decreases, leading to a finite value for $\kappa'_{equil}(\mu_{crit})$.

Interestingly, for $l = \infty$, the derivative at the error catastrophe becomes infinite as $k \to \infty$ for $\lambda > 0$.

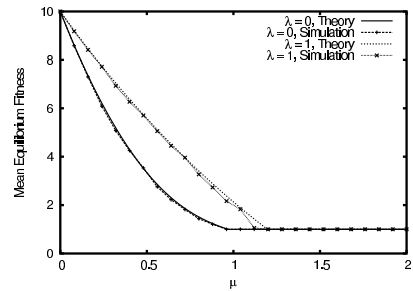

FIG. 1: Comparison of theory and simulation results for $l = 0$.

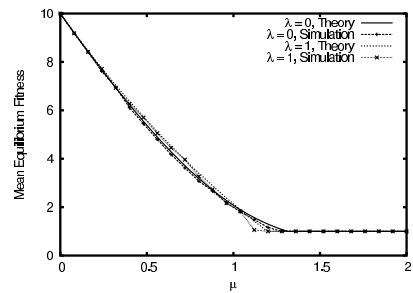

FIG. 2: Comparison of theory and simulation results for $l = 1$.

### C. Stochastic simulations

In order to compare the results of our theory with actual numerics, we ran stochastic simulations of finite populations of replicating organisms. Specifically, we determined $\bar{\kappa}(t = \infty)$ at various values of $\mu$ for $l = 0$ (Figure 1), $l = 1$ (Figure 2), and $l = \infty$ (Figure 3). We considered genomes of length 40, and populations of 1,000 organisms. Our results were obtained by averaging over 10 independent runs, where each run consisted of 10,000 time steps of size 0.01. We took $k = 10$.

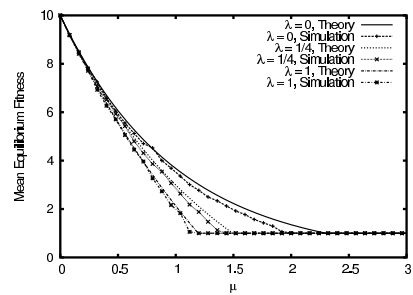

FIG. 3: Comparison of theory and simulation results for $l = \infty$.

Note the excellent agreement between theory and simulation. One interesting feature to note is that for $l = 1$, the $\lambda = 1$ fitness is slightly larger than the $\lambda = 0$ fitness for almost all $\mu$ below the error catastrophe. However, the $\lambda = 1$ error catastrophe happens before the $\lambda = 0$ error catastrophe, consequently, there is a region where the $\lambda = 0$ fitness becomes greater. We give a possible explanation for this phenomenon: Below the error catastrophe, it is advantageous to maintain the highest replication fidelity possible, which is done by maximizing the lesion repair efficiency. A tolerance of one lesion is not sufficient to provide a selective advantage for inefficient lesion repair. However, when the $\lambda = 1$ error catastrophe is reached, at $\mu_{crit} \approx 2\ln 2$, then lesion repair no longer reduces the error rate by a sufficient amount to avoid the death of the population. At this point, it becomes advantageous to turn lesion repair off. With lesion repair turned off, any replication mistakes that are made remain in the daugher strand. Thus, the parent strand is preserved, and since the master genome can tolerate some lesions, it is still possible to produce a viable genome. On a subsequent replication cycle, the unchanged parent strands can be reintegrated into a master genome.

As described above, the result of these competing effects is that the $\lambda = 1$ fitness is greater than the $\lambda = 0$ fitness almost until the $\lambda = 1$ error catastrophe. However, just before the $\lambda = 1$ error catastrophe, the mean fitness switches, and the $\lambda = 0$ catastrophe happens after the $\lambda = 1$ catastrophe.

We should note that at $\mu_{crit} \approx 2\ln 2 = \ln 4$, an average of about 1 mismatch is made per daughter strand synthesis. Thus, without lesion repair, the tolerance of a one-base lesion in the master genome is just sufficient to preserve viability at the $\lambda = 1$ value for $\mu_{crit}$. As the tolerance for lesions grows, the selective advantage for turning off lesion repair even below the error catastrophe increases as well. Eventually, for $l = \infty$, when lesion repair is turned off we obtain conservative replication, which has a higher fitness than semiconservative replication at all mutation rates.

## VI. CONCLUSIONS AND FUTURE RESEARCH

This paper developed the quasispecies equations suitable for describing semiconservative replication with imperfect lesion repair. The work presented here may be regarded as a continuation of the work in [27], which provided the quasispecies equations for semiconservative replication, under the assumption of perfect lesion repair. We solved the model for a genealized "single-fitness-peak" landscape where the master genome can sustain a finite number $l$ of lesions and remain viable. For future research, it will be interesting to consider the behavior of the model for more realistic landscapes. Specifically, we would like to explore the behavior of the model when a genome is viable even for positive values of $l_C$. In the original semiconservative quasispecies equations, a fitness landscape which allows for a finite number of point mutations before loss of viability does not delay the occurrence of the error catastrophe beyond what is predicted in the single-fitness-peak model [22]. We expect this result to change when lesion repair is imperfect.

Furthermore, we plan to apply the quaispecies model with imperfect lesion repair to stem cell growth, explicitly incorporating the "immortal strand" segregation mechanism. Due to the nature of stem cell division and tissue development, such a model moves beyond the simple model of genomes replicating in a chemostat [22, 38], where each genome has an equal probability of being removed from the population. Indeed, we will need to develop a further extension of the imperfect lesion repair quasispecies equations, using techniques from what is known as *evolutionary graph theory*.

### Acknowledgments

This research was supported by the National Institutes of Health. The authors would like to thank Yisroel Brumer and Eric J. Deeds for helpful conversations.

## APPENDIX A: DETERMINATION OF $\lim_{\mu \to \mu_{crit}^-} \kappa_{equil}(\mu)$

To evaluate $\kappa'_{equil}(\mu)$ for arbitrary lesion repair, we start with the fact that for $\mu < \mu_{crit}$, $\kappa_{equil}(\mu)$ satisfies,

$$0 = \kappa_{equil}(\mu)^2 - A(\mu, \lambda)\kappa_{equil}(\mu) - B(\mu, \lambda) \quad (A1)$$

Differentiating both sides gives,

$$0 = 2\kappa_{equil}\kappa'_{equil} - \partial_\mu A \kappa_{equil} - A\kappa'_{equil} - \partial_\mu B \quad (A2)$$

When $\mu = \mu_{crit}$ we have $\kappa_{equil} = 1$, giving,

$$\kappa'_{equil} = -k\frac{e^{-\mu(1-\frac{\lambda}{2})}((1-\frac{\lambda}{2})(2+f_l(\mu,\lambda)) - (1-\lambda)f_{l-1}(\mu,\lambda)) + \frac{\lambda}{2}e^{-\mu\frac{\lambda}{2}}}{3 + f_l(\mu,\lambda)e^{-\mu(1-\frac{\lambda}{2})} - e^{-\mu\frac{\lambda}{2}} - k((1+f_l(\mu,\lambda))e^{-\mu(1-\frac{\lambda}{2})} - 1)}$$
$$+ \frac{((1-\frac{\lambda}{2})f_l(\mu,\lambda) - (1-\lambda)f_l(\mu,\lambda))e^{-\mu(1-\frac{\lambda}{2})} - \frac{\lambda}{2}e^{-\mu\frac{\lambda}{2}}}{3 + f_l(\mu,\lambda)e^{-\mu(1-\frac{\lambda}{2})} - e^{-\mu\frac{\lambda}{2}} - k((1+f_l(\mu,\lambda))e^{-\mu(1-\frac{\lambda}{2})} - 1)} \quad (A3)$$

where we used the identity $\partial_\mu f_l(\mu,\lambda) = (1-\lambda)f_{l-1}(\mu,\lambda)$.

Note that the numerator of the first fraction is positive for $\mu > 0$ (where we are neglecting the factor of $-k$), so as $k \to \infty$ and $\mu \to \mu_{crit}$, if $(1 + f_l(\mu_{crit},\lambda))e^{-\mu_{crit}(1-\frac{\lambda}{2})} - 1 = 0$, then $\kappa'_{equil} \to -\infty$. Conversely, if $(1+f_l(\mu_{crit},\lambda))e^{-\mu_{crit}(1-\lambda/2)} - 1 \neq 0$, then the denominator ensures that the derivative remains finite.

Now, at the error catastrophe, we have $A(\mu_{crit},\lambda) = 1 - B(\mu_{crit},\lambda)$. Therefore, plugging into our expression for $\kappa'_{equil}$, we get that $\kappa'_{equil}$ is infinite if and only if $e^{-\mu_{crit}\lambda/2} + e^{-\mu_{crit}(1-\lambda/2)} - 1 = 0$. However, we have shown that $\kappa'_{equil}$ is infinite if and only if $(1+f_l(\mu_{crit},\lambda))e^{-\mu_{crit}(1-\lambda/2)} - 1 = 0$. Therefore, if $\kappa'_{equil}$ is infinite, then we must have $f_l(\mu_{crit},\lambda) = e^{\mu_{crit}(1-\lambda)} = f_\infty(\mu_{crit},\lambda)$. For finite $l$, note that $f_l(\mu_{crit},\lambda) \leq f_\infty(\mu_{crit},\lambda)$, with equality only when $\mu_{crit}(1-\lambda) = 0 \Rightarrow \lambda = 1$.

Therefore, for *finite* $l$, $\lim_{\mu \to \mu_{crit}} \kappa'_{equil}(\mu)$ remains finite as $k \to \infty$ as long as $\lambda < 1$.

When $l = \infty$, then $\kappa_{equil}(\mu) = k(e^{-\mu\lambda/2} + e^{-\mu(1-\lambda/2)} - 1)$ below the error catastrophe. It is readily shown that, except for $\lambda = 0$, the derivative at the error catastrophe becomes infinite as $k \to \infty$.

### APPENDIX B: NOTES ON THE IMPLEMENTATION OF THE STOCHASTIC SIMULATIONS

Stochastic simulations are run using a finite population of $N$ replicating genomes of length $L$. The simulation is run out to some prespecified time $T$ at time steps of some prespecified $\Delta t$. We try to choose $T$ large enough to obtain good equilibration of the population, and $\Delta t$ small enough so that one can reasonably make a continuous time assumption.

At each time step, we cycle over each organism in the population, and determine whether it replicates in that time interval. The replication probability $p_{\{\sigma,\sigma'\}}$ of an organism with genome $\{\sigma,\sigma'\}$ may be computed from the first-order growth rate constant in one of two ways: $p_{\{\sigma,\sigma'\}} = \min\{\kappa_{\{\sigma,\sigma'\}}\Delta t, 1\}$, or $p_{\{\sigma,\sigma'\}} = 1 - e^{-\kappa_{\{\sigma,\sigma'\}}\Delta t}$. In practice, we choose $\Delta t$ to be sufficiently small so that the two definitions yield almost identical results.

If an organism replicates, then it is effectively destroyed, and it produces two new organisms. At the end of each replication cycle, we randomly remove organisms from the population until the population size returns to $N$.

---